\documentclass[preprint,double-spaced,superscriptaddress,showpacs,prb]{revtex4}
\usepackage{amssymb}
\usepackage{graphicx}
\usepackage{dcolumn}
\usepackage{bm}

\begin{document}
\title{Exchange and correlation effects on the plasmon dispersions and the
Coulomb drag in low-density electron bi-layers}
\author{S. M. Badalyan}\email{smbadalyan@ysu.am}
\affiliation{Radiophysics Department, Yerevan State University, 1
A. Manoukian St., Yerevan, 375025 Armenia}\affiliation{Department
of Physics and Institute for Condensed Matter Theory, Chonnam
National University, Gwangju 500-757, Korea}
\author{C. S. Kim}
\affiliation{Department of Physics and Institute for Condensed
Matter Theory, Chonnam National University, Gwangju 500-757,
Korea}
\author{G. Vignale}
\affiliation{Department of Physics and Astronomy, University of
Missouri - Columbia, Missouri 65211, USA}
\author{G. Senatore}
\affiliation{Dipartimento di Fisica Teorica, Universit\`a di
Trieste, Strada Costiera 11, 34014 Trieste, Italy} \affiliation{
INFM-CNR Democritos National Simulation Center, Trieste, Italy}

\date{\today}

\begin{abstract}
We investigate the effect of exchange and correlation (xc)  on the
plasmon spectrum and the Coulomb drag between spatially separated
low-density two-dimensional electron layers. We adopt a new
approach, which employs dynamic xc kernels in the calculation of
the bi-layer plasmon spectra and of the plasmon-mediated drag, and
static many-body local field factors in the calculation of the
particle-hole contribution to the drag. The spectrum of bi-layer
plasmons and the drag resistivity are calculated in a broad range
of temperatures taking into account both intra- and inter-layer
correlation effects.  We observe that both plasmon modes are
strongly affected by xc corrections. After the inclusion of the
complex dynamic xc kernels, a decrease of the electron density
induces shifts of the plasmon branches in opposite directions. And
this is in stark contrast to the tendency obtained within the RPA
that both optical and acoustical plasmons move away from the
boundary of the particle-hole continuum with a decrease in the
electron density. We find that the introduction of xc corrections
results in a significant enhancement of the transresistivity and
qualitative changes in its temperature dependence. In particular,
the large high-temperature plasmon peak that is present in the
random phase approximation is found to disappear when the xc
corrections are included. Our numerical results at low
temperatures are in good agreement with the results of recent
experiments by M. Kellogg {\it et al.}, Solid State Commun.
\textbf{123}, 515 (2002).

\end{abstract}

\pacs{71.45.Gm, 73.20.Mf, 73.63.Hs}
\maketitle



%






\section{Introduction}

In a double layer two-dimensional electron system (2DES) the
additional degree of freedom associated with the layer index plays
a key role in many recently discovered collective phenomena
\cite{eisen}. Suffice it to mention the freshly provided evidence
that in the quantum Hall regime, near the half filling of
individual layers, strong electron-hole inter-layer correlations
cause Bose-Einstein condensation of excitons to occur in bi-layer
2D electron and hole systems \cite{bec}. In these structures, the
ability to make separate electrical contacts to each layer has
allowed the experimental detection of transport of neutral
excitons as a counter-flowing current of electrons and holes in
two layers. The independent control of layers is also critical for
the experimental realization of the frictional drag
\cite{pogreb,price} which manifests itself under the condition
that the second layer is an open circuit. A current driven along
the first layer induces, via momentum transfer, a voltage in the
second layer, which is measured experimentally. The drag effect
complements and enriches the traditional experimental methods and
in the last decade has proved to be a powerful tool to probe
inter-layer electron-electron (e-e) interaction\cite{rojo}.

Recently the drag measurements have been extended to the limit of
very low carrier concentrations. The dimensionless parameter
$r_{s}=\sqrt{2} /(k_{F}a_{B}^{\ast })$, which describes the
carrier density, $n$, and measures the strength of the
electron-electron interaction \cite{GV05}, varies approximately
from $10$ to $20$ in the experiment on hole samples by
Pillarisetty \textit{et al}. \cite{pillar,pillar2}. Here
$k_{F}=\sqrt{2\pi n} $ is the Fermi wave vector, and $a_{B}^{\ast
}=\hbar ^{2}\kappa _{0}/m^{\ast}e^{2}$ is the effective Bohr
radius, with $\kappa _{0}$ the static dielectric constant and
$m^*$ the effective mass of the carriers; $a_{B}^{\ast }=9.79 $
and $3.45$ nm, respectively, in the conduction ($m^*_e=0.067m_0$)
and valence ($m_h^*=0.19m_0$) bands of GaAs. In the experiment on
electron samples by Kellogg \textit{et al.}\cite{kellogg}, $r_{s}$
varies approximately from $2.5 $ to $4.5$ and is appreciably
smaller, owing to the small effective mass of the electrons. From
a dimensional analysis it is clear, however, that in both types of
samples, the Coulomb potential energy dominates the kinetic
energy, and an adequate description of the drag cannot be provided
by simple theories, which do not include the strong
exchange-correlation (xc) effects. Consistent with this
expectation, both experimental groups
\cite{kellogg,pillar,pillar2} have found that the measured drag
rate far exceeds (by more than a factor 200 in the hole samples)
the value obtained from simple Boltzmann-equation
calculations\cite{jauho}. In addition, systematic deviations are
observed from the quadratic temperature dependence of the drag
rate (the characteristic Fermi-liquid behavior of the Coulomb drag
at low temperatures).

One might hope that the above features could be understood in the
framework of the phonon-mediated drag mechanism
\cite{gramila2,rubel,noh2,joerger,tso1,bonsager,smb1}. However,
the phonon-mediated drag provides a dominant mechanism only in
samples with large inter-layer spacing where the absolute value of
the measured drag rate is much smaller than that observed in these
experiments. Moreover, the phonon mechanism is not supported by
the measured density-ratio-dependence of drag, which shows no peak
at matched carrier densities in two layers, typical for that
processes. Thus, both experimental groups
\cite{kellogg,pillar,pillar2} conclude that in these dilute 2DES
the strong Coulomb interaction effects are mainly responsible for
the observed new features.

The carrier interaction effects on drag have been addressed
previously in several experimental~\cite{sivan,hill,noh1} and
theoretical \cite{tso2,flensberg,swierkowski} papers. While the
theoretical prediction by Flensberg and Hu~\cite{flensberg} of a
strong enhancement of drag by plasmons has been experimentally
verified\cite{hill} in high density electron samples, important
differences have been reported\cite{noh1} from the results
obtained within the random phase approximation (RPA).

Lately, motivated by the recent set of low-density experiments
\cite{kellogg,pillar,pillar2}, the drag resistivity has been
calculated by Hwang \textit{et al.}~\cite{hwang} and Yurtserver
\textit{et al.}~\cite{yurt}. Both works have included only the
exchange interaction effects in the static limit via $q$-dependent
but $\omega $-independent local field factors (LFF). Besides, in
the adopted approximation the nondiagonal inter-layer LFF have
been taken to be zero\cite{hwang}, while for the intra-layer LFF
the simple Hubbard approximation has been used, which
significantly underestimates the LFF. As in the low density regime
the inter-particle spacing in each layer becomes comparable with
the inter-layer separation, one should expect that inter-layer
correlations play an important role. At relatively high
temperatures the dynamic xc effects are also critical. They can be
especially important in the hole samples where, even at
sufficiently low temperatures, the samples are effectively in the
high temperature regime because of the small Fermi energy.

In this paper we investigate the xc effects on the Coulomb drag in
low-density e-e bi-layers. The spectrum of plasmons of the coupled
2DES and the temperature dependence of the drag resistivity are
calculated in a broad range of temperatures. Our calculations
include both intra- and inter-layer xc effects. We propose a new
approach, which employs dynamic xc kernels in the calculation of
the bi-layer plasmon spectra and of the plasmon-mediated drag,
while the particle-hole contribution to the drag is still
calculated by means of static many-body LFF. We find that the
introduction of xc corrections results in a significant
enhancement of the drag resistivity and qualitative changes in its
temperature dependence. In particular, a large high-temperature
plasmon peak that is present in the RPA disappears when the xc
corrections are included. At low temperatures our numerical
results are in good agreement with the experimental findings by
Kellogg \textit{et al.}~\cite{kellogg}.

The paper is organized as follows. In Section II we outline the
theoretical formulation of the frictional drag and provide the
main formulas for its calculation beyond the RPA. In Section III a
method is developed to relate the spin-averaged dynamic xc kernels
and static many-body LFF of a bi-layer 2DES  to the spin-resolved
xc kernels and LFF of a single electron gas layer of the same
total density. The results of our calculations are given in
Sections IV-VI. First we present the spectra of bi-layer plasmons
for different densities and discuss the effect of dynamic xc on
the dispersion of each plasmon mode, comparing the dispersion
curves with the corresponding curves obtained within the RPA. In
Section V we study the temperature dependence of the drag
resistivity at low densities within the RPA. Both plasmon and
particle-hole contributions to the drag are considered. In the
next Section we investigate the many-body xc effects on the drag
resistivity for different densities and examine the distinctive
features of the drag temperature dependence in comparison with the
RPA. In the same Section VI we compare the obtained theoretical
results with the experimental findings in low density e-e
bi-layers. Finally, Section VII summarizes the main results of the
paper.

\section{Theoretical formulation}

We calculate the temperature dependence of the drag resistivity
starting from the formula, obtained in the Kubo formalism
\cite{oppen}
\begin{eqnarray}
\rho _{_{D}} &=&{\frac{\hbar ^{2}}{2e^{2}\sigma _{1}\sigma _{2}TA}\frac{
d\sigma _{1}}{dn_{1}}\frac{d\sigma _{2}}{dn_{2}}}\sum_{\overrightarrow{q}%
}q^{2}\int_{0}^{\infty }\frac{d\omega }{2\pi }\left\vert W_{12}(q,\omega
)\right\vert ^{2}  \nonumber \\
&&\times \frac{\text{Im}\Pi _{1}^{0}(q,\omega )\text{Im}\Pi
_{2}^{0}(q,\omega )}{\sinh ^{2}(\hbar \omega /2T)}~,  \label{eq1}
\end{eqnarray}
where $A$ is the normalization area, $\hbar \omega $ and $\hbar
\overrightarrow{q}$ the transferred energy and in-plane momentum
from the layer $1$ to the layer $2$, $W_{12}(q,\omega )$ the
dynamically screened inter-layer e-e interaction including the
intra- and inter-layer many-body xc effects of a bi-layer 2DES,
$\Pi _{1,2}^{0}(q,\omega )$ the finite temperature electron
polarization function of an individual 2DES in the absence of
inter-particle Coulomb interaction. The imaginary and real parts
of $\Pi _{1,2}^{0}(q,\omega )$ are obtained from the Maldague
formula \cite{maldague,flensberg} and are given by
\begin{eqnarray}
\text{Im}\Pi ^{0}(q,\omega ) &=&g_0~\frac{\sqrt{\pi t}}{4x}\left[
F_{-1/2}\left(
\frac{\mu (t)-\zeta _{+}^{2}(x,y)}{t}\right) \right.  \nonumber \\
&&-\left. F_{-1/2}\left( \frac{\mu (t)-\zeta
_{-}^{2}(x,y)}{t}\right) \right]~,
\end{eqnarray}
and
\begin{eqnarray}
\text{Re}\Pi ^{0}(q,\omega ) &=&g_0~ \left\{
1-e^{-1/t}-\frac{1}{2x}\left[ M\left(
t,\zeta _{+}^{2}(x,y)\right) \right.\right.  \nonumber \\
&&-\left.\left. \text{sign}\left( \zeta _{-}(x,y)\right) \text{
}M\left( t,\zeta _{-}^{2}(x,y)\right) \right]\right\}~,
\end{eqnarray}
where we have introduced the following dimensionless quantities:
$t=T/\varepsilon _{F},$ $x=q/2k_{F}$, $y=\omega /4\varepsilon
_{F}$, $\mu (t)=t\ln \left( e^{1/t}-1\right)$, $\zeta _{\pm
}(x,y)=y/x\pm x$ with $ \varepsilon _{F}$ and $k_{F}$ being the
Fermi energy and the Fermi wave vector. The function

\begin{equation}
F_{-1/2}(u)=1/\sqrt{\pi }\int_{0}^{\infty }dz/\sqrt{z}\left( \exp
(z-u)+1\right)
\end{equation}
is the Fermi integral, the function
\begin{equation}
M\left( t,u\right) =1/4t\int_{0}^{u}dz\sqrt{u-z}/\cosh \left[
\left( z-\mu (t)\right) /2t\right] ^{2}~,
\end{equation}
and $g_{0}=m^{\ast }/\pi \hbar ^{2}$ is the density of states of
the noninteracting 2DES. Further we assume that in Eq.~(\ref{eq1})
the layer conductivities $\sigma _{1,2}$\ depend linearly on the
electron densities $n_{1,2}$. In general, this is an acceptable
approximation for the conducting 2DES in the magnetic-field-free
case. Deviations from the linear regime can occur at very low
electron densities. Here we  consider only  the regime
corresponding to $r_{s}\leq 5$;  in this regime we expect the
neglect of the corrections to the $\sigma_{1,2}\varpropto n_{1,2}$
to be a reasonable approximation.

Our starting equation (\ref{eq1}) is an approximation, strictly
justifiable only to second order in the Coulomb interaction. The
correct expression for the transresistivity involves a force-force
response function, which is a four-point response function and
cannot be reduced, in general, to a product of  two-point response
functions.
In order to include the xc effects on the dynamically screened
interlayer Coulomb interaction, in our calculations we have
exploited an approximation scheme for $W_{12}$ proposed by Zheng
and MacDonald \cite{zheng2} (see also Ref.~
\onlinecite{swierkowski}). Within this scheme an exact matrix
Dyson equation for the Coulomb propagator \cite{mahan},
${\hat{W}}(q,\omega )$, is approximated by
\begin{equation}
{\hat{W}}(q,\omega )={\hat{V}}_{\text{eff}}(q,\omega
)-{\hat{V}}_{\text{eff} }(q,\omega ){\hat{\Pi}}(q,\omega
){\hat{V}}_{\text{eff}}(q,\omega ) \label{dyson}
\end{equation}%
with the full polarization function, ${\hat{\Pi}}(q,\omega )$,
defined in
terms of the non-interacting polarization function, ${\hat{\Pi}}%
_{0}(q,\omega )$, and unscreened effective Coulomb interactions,
${\hat{V}}_{ \text{eff}}\left( q,\omega \right) $, as
\begin{equation}
{\hat{\Pi}}(q,\omega )=\frac{{\hat{\Pi}}_{0}(q,\omega
)}{{\,}1+{\hat{V}}_{ \text{eff}}(q,\omega
){\hat{\Pi}}_{0}(q,\omega )}~. \label{polar}
\end{equation}%
Here all quantities are $2\times 2$ matrices.  The effective
Coulomb interactions ${\hat{V}}_{\text{eff}}$ (whose explicit form
is given below) build in Eqs.~\ref{dyson} and~\ref{polar} the
vertex corrections as  sums of ladder diagrams evaluated within a
"local approximation" \cite{vigs}. Thus, the non-diagonal
interaction matrix element $W_{12}(q,\omega )$ is given by
\begin{equation}
W_{12}(q,\omega )=\frac{V_{\text{eff},12}(q,\omega )}{\varepsilon
_{\text{bi} }(q,\omega )}  \label{sec}
\end{equation}%
with the bi-layer screening function
\begin{equation}
\varepsilon _{\text{bi}}(q,\omega )=\varepsilon _{1}(q,\omega
)\varepsilon _{2}(q,\omega )-V_{\text{eff},12}(q,\omega )^{2}\Pi
_{1}^{0}(q,\omega )\Pi _{2}^{0}(q,\omega )~.
\end{equation}%
Here we introduce the screening functions of individual layers
\begin{equation}
\varepsilon _{1,2}(q,\omega )=1+V_{\text{eff},11,22}(q,\omega )\Pi
_{1,2}^{0}(q,\omega )~.  \label{singscreening}
\end{equation}%
The approximation given by
Eqs.~(\ref{dyson})-(\ref{singscreening}) accounts for the effect
of dynamical screening of the Coulomb potential via the bi-layer
screening function, $\varepsilon _{\text{bi}}(q,\omega )$, and
includes the correlations between the test electron and the
induced charge via the effective Coulomb interactions,
${\hat{V}}_{\text{eff}}\left( q,\omega \right) $. In this
approximation xc LFF manifest themselves only via the intra- and
inter-layer unscreened effective interactions
\begin{equation}
V_{\text{eff},ij}(q,\omega )=v(q)\left(
1-G_{\text{xc,}ij}(q,\omega )\right) F_{ij}(qd,q\Lambda )
\label{veff}
\end{equation}%
where the intra- and inter-layer LFF $G_{\text{xc},ij}(q,\omega )$
decrease effectively the bare Coulomb interaction, $v(q)=2\pi
e^{2}/\kappa _{0}q$, by a factor of $1-G_{\text{xc,}ij}(q,\omega
)$ ($i,j=1,2$ are the layer indices). Notice that the RPA is
recovered if all the LFF in Eq.~\ref{sec} are set to zero.

Further we assume that the electrons in each layer are confined in
a square well potential of infinite height and that scattering
processes take place only within the lowest subband of each well.
Assuming $\rho _{1}(z)=(2/d)\,\sin (\pi z/d)^{2}$ and $\rho
_{2}(z)=\rho _{1}(z+\Lambda )$ as explicit forms for the electron
density profiles in layers $1$ and $2$, respectively, one obtains
\cite{jauho} for the corresponding intra- and inter-layer form
factors (see   Eq.~\ref{veff})
\begin{equation}
F_{11}(\eta ,\xi )=\frac{8\pi ^{2}+3\eta ^{2}}{\eta \left( 4\pi
^{2}+\eta ^{2}\right) }-\frac{32\pi ^{4}(1-e^{-\eta })}{\eta
^{2}\left( 4\pi ^{2}+\eta ^{2}\right) ^{2}}
\end{equation}%
and%
\begin{equation}
F_{12}(\eta ,\xi )=\frac{64\pi ^{4}\sinh \left( \eta /2\right)
^{2}}{\eta ^{2}\left( 4\pi ^{2}+\eta ^{2}\right) ^{2}}e^{-\xi }~,
\end{equation}%
where $\Lambda $ is the center-to-center inter-layer separation
and $d$ the width of quantum wells.

\section{Dynamic xc kernel and static LFF}

In general the local field factors of a bi-layer system,
$\widetilde{{G}}_{ij}^{\mu \nu }(q,\omega )$, depend on four
indices: the spin indices $\mu$ and $\nu$,  taking values
$\uparrow$ and $\downarrow $,  and the layer indices  $i$ and $j$
taking values $1$ and $2$.  But the LFFs that appear in
Eq.~(\ref{veff}),  and hence in the effective interaction
$W_{12}$,  are averaged over the spin indices, at least as long as
we assume  that our bi-layer is {\it not} spin-polarized.    Thus
we have
\begin{eqnarray}
{G}_{\text{xc},ij}(q,\omega ) &\equiv&\frac{1}{4} \sum_{\mu\nu}
\widetilde{{G}} _{ij}^{\mu\nu}(q,\omega)~. \label{avlff4}
\end{eqnarray}

Unfortunately, reliable calculations of the LFFs for bi-layers
(for example by the diffusion Monte Carlo method) have not been
done to date.  In order to proceed in the absence of this crucial
information we propose an alternative approach, in which the
spin-averaged LFFs of the bi-layer are expressed in terms of the
spin-resolved LFFs of a single electron layer of the same total
density $n=n_1+n_2$ -- the {\it ``reference monolayer"}. We
suggest to exploit the analogy between the spin index of the
reference monolayer and the pseudospin (layer) index of the
original bi-layer, neglecting the spin index in the bilayer
altogether. We thus map the layer index (in the bilayer) onto the
spin index in the monolayer and represent the intra- and
inter-layer LFFs of the bi-layer  in terms of the "spin-channel"
and "charge-channel" LFFs of the reference monolayer, denoted
respectively by $G_{\text{m}}^{-}(q,\omega )$ and
$G_{\text{m}}^{+}(q,\omega )$ in the following manner:
\begin{eqnarray}
G_{\text{xc},{11}}(q,\omega ) &=&G_{\text{m}}^{+}(q,\omega )+G_{\text{m}%
}^{-}(q,\omega )~, \nonumber  \\
G_{\text{xc},{12}}(q,\omega ) &=&G_{\text{m}}^{+}(q,\omega )-G_{\text{m}%
}^{-}(q,\omega )~.  \label{bilff}
\end{eqnarray}%
Of course this simple representation is possible only in the
symmetric case, $n_{1}=n_{2}$, when we have
${G}_{\text{xc},11}(q,\omega )={G}_{\text{xc} ,22}(q,\omega )$ and
${G}_{\text{xc},12}(q,\omega )={G}_{\text{xc} ,21}(q,\omega )$.

It is evident that Eq.~(\ref{bilff}) is not exact: for one thing,
the interaction between electrons in different layers is different
from the interaction between two electron in the same layer -- a
difference that does not exist between parallel and
antiparallel-spin electrons in the reference monolayer.  Subtle
differences persist even in the limit of zero inter-layer
separation.  In the reference monolayer the Pauli exclusion
principle  keeps two electrons of the same spin orientation from
coming close to each other;  but in the bi-layer  two electrons in
the same layer (i.e. with parallel isospin) can come together
provided their real spins are opposite.

\begin{figure}[h] \centering
\includegraphics[width=10cm,angle=-90]{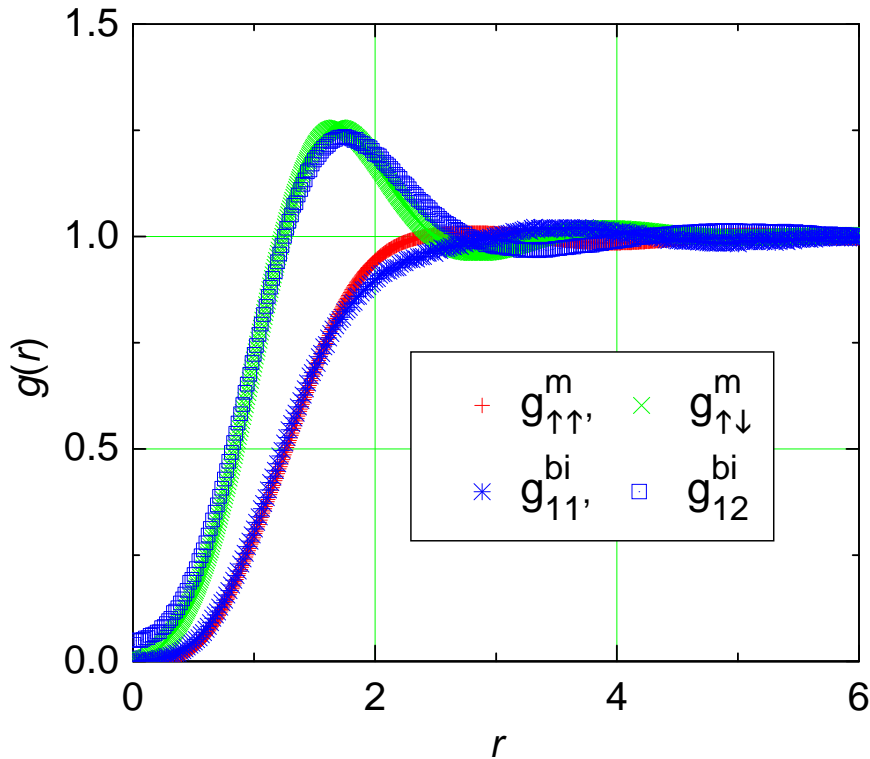}
\caption{Bi-layer intra- and inter-layer pair-correlation
functions $g_{11}^{bi}(r)$ and $g_{12}^{bi}(r)$ and spin resolved
pair-correlation functions $g_{\uparrow\uparrow}^{m}(r)$ and
$g_{\uparrow \downarrow}^{m}(r)$ for a monolayer as directly
obtained from the quantum Monte Carlo simulations, respectively,
from Refs.~\onlinecite{bilayergr} and \onlinecite{french}. The
total $r_s=7.07$ both in the monolayer and bi-layer. It
corresponds to the in-layer $r^{in-layer}_s=10$ of the bi-layer.
The inter-layer spacing of the bi-layer is $\Lambda =0.2
r^{in-layer}_s a^*_B$.} \label{dmc}
\end{figure}

In spite of these difficulties we find that the
approximation~(\ref{bilff}) is quite reasonable at small
inter-layer separations.   We get a good idea of the electronic
correlations in a system by examining its pair-distribution
functions.  In Fig.~\ref{dmc} we compare  the spin-averaged
pair-distribution functions of a bi-layer, $g_{11}^{bi}(r)$
(intra-layer) and $g_{12}^{bi}(r)$ (inter-layer),  as  obtained
from the diffusion Monte Carlo simulations\cite{bilayergr}, with
the spin-resolved pair-distribution functions $g_{\uparrow
\uparrow }^{m}(r)$ and $g_{\uparrow \downarrow }^{m}(r)$ of the
reference monolayer, \cite{french}. It is seen that there is an
overall good agreement between the corresponding distribution
functions, i.e. $g_{11}^{bi}$ agrees with $g_{\uparrow \uparrow
}^{m}$ and $g_{12}^{bi}$ with $g_{\uparrow \downarrow }^{m}$. This
agreement provides much support for the use of Eq.~\ref{bilff}. As
$\Lambda$ increases, the agreement between the bi-layer
pair-distribution function $g_{12}^{bi}(r)$ and the monolayer
pair-distribution function $g_{\uparrow\downarrow}^{m}(r)$ will
deteriorate.  However,  the disagreement between the interlayer
correlations and the unlike spin correlation in the reference
monolayer is more severe at small $r$ than at large
$r$\cite{bilayergr}.  The pair-distribution function at small
values of $r$, in turn, should control the behavior of the LFF at
large wave vector $q\gg2k_F$, while the main contribution to the
Coulomb drag, at low densities, come from scattering processes
with $q$ near $2k_F$.  Thus, we see that our approach could be
justified for the problem at hand, even when the inter-layer
spacing is less than or comparable to the  in-layer inter-particle
distance, $\Lambda\lesssim r_s a^*_B$, which is the case in the
experiments of interest here.


Let us now come to the discussion of the LFFs in the reference
monolayer.  Here we have to introduce further approximations. When
treating the plasmon contribution to the Coulomb drag in our
calculations we approximate the dynamic xc kernels,
$f_{\text{xc},ij}(q,\omega)\equiv-v(q)G_{\text{xc},ij}(q,\omega)$
by their long-wavelength limits and employ the dynamic spin-spin
and charge-charge xc kernels, evaluated by Qian and Vignale in
Refs.~\onlinecite{qv} and \onlinecite{qian}. In this way we
include the frequency dependence of the dynamic charge-charge LFF,
while their wave vector dependence is implicitly assumed to be
linear so that the charge-charge xc kernel is independent of $q$.
This is a reasonable approximation in view of the nearly linear
behavior of the exact static LFF versus $q$ in the single
layer\cite{GV05}. On the other hand, the spin-spin dynamic xc
kernel is strongly dependent on $q$ and diverges as $q^{-2}$ at
small wave vectors\cite{qcv}. Thus, the intra- and inter-layer
dynamic xc kernels in the long wavelength limit are represented as
\begin{equation}
f_{\text{xc,}ij}(q,\omega )=B^{+}(\omega )+\left( 2\delta
_{ij}-1\right) \left[ \frac{A(\omega )}{q^{2}}+B^{-}(\omega
)\right]  \label{fxc}
\end{equation}
where $A(\omega )$ and $B^{\pm }(\omega )$ are finite functions of
$\omega$ and have been calculated to the leading order in the
Coulomb interaction in Refs.~\onlinecite{qv} and
\onlinecite{qian}. The authors of this paper have evaluated the
four-vertex diagrams in the expansion of the response function of
individual 2DES in the low-frequency/high-density limit. The
interpolation formulas for Im~$A(\omega )$ and Im~$B^{\pm }(\omega
)$ have been constructed by combining the results for the
imaginary parts with sum rules and previously known exact
behaviors in the high-frequency limit \cite {hfl}. Thus, having
the imaginary parts of $A(\omega )$ and $B^{\pm }(\omega )$, we
obtain the corresponding real parts from the Kramers-Kroning
relations.

\begin{figure*}[h,t,b,p]
\centering
\includegraphics[width=6cm,angle=-90]{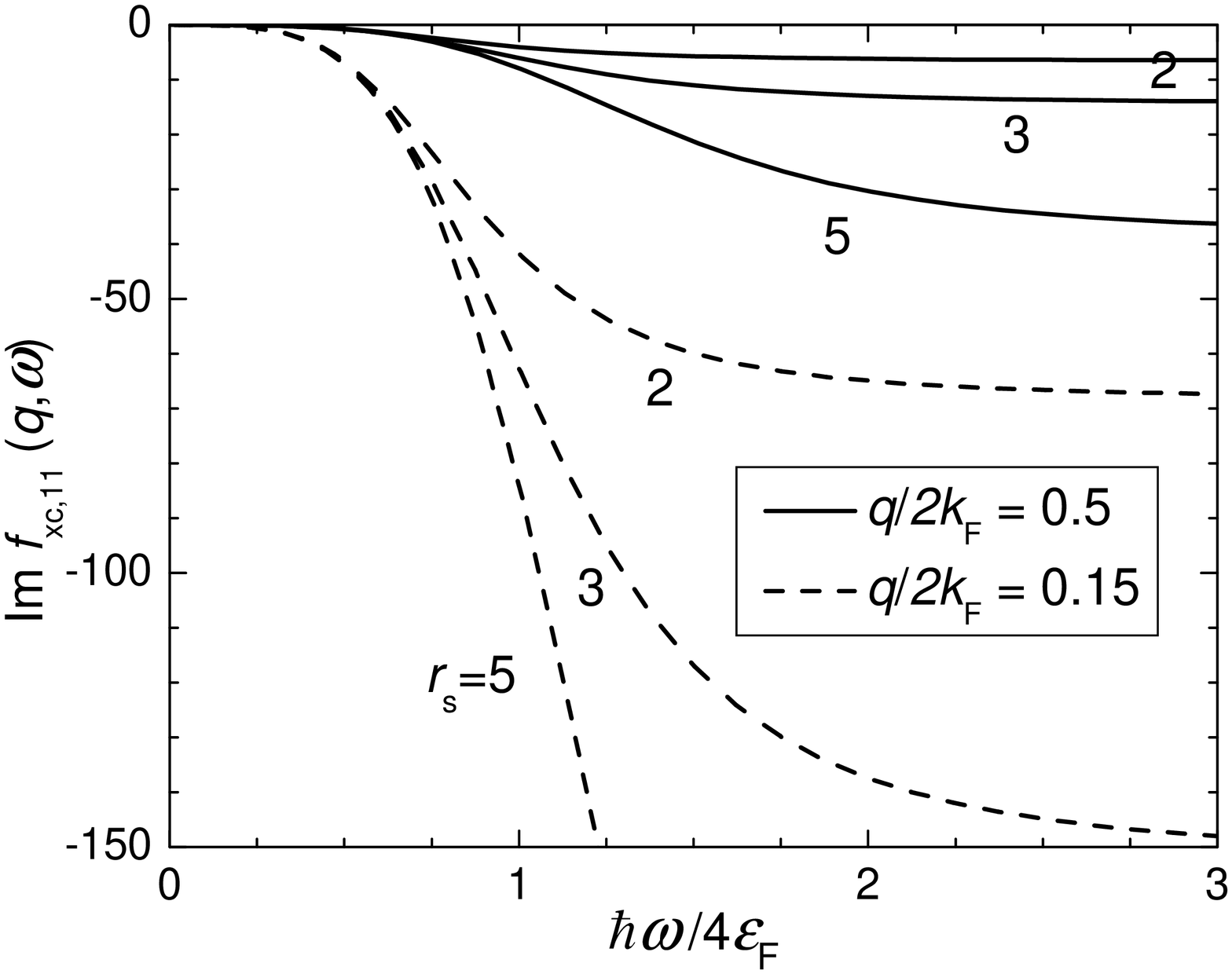} \hspace{1cm} %
\includegraphics[width=6cm,angle=-90]{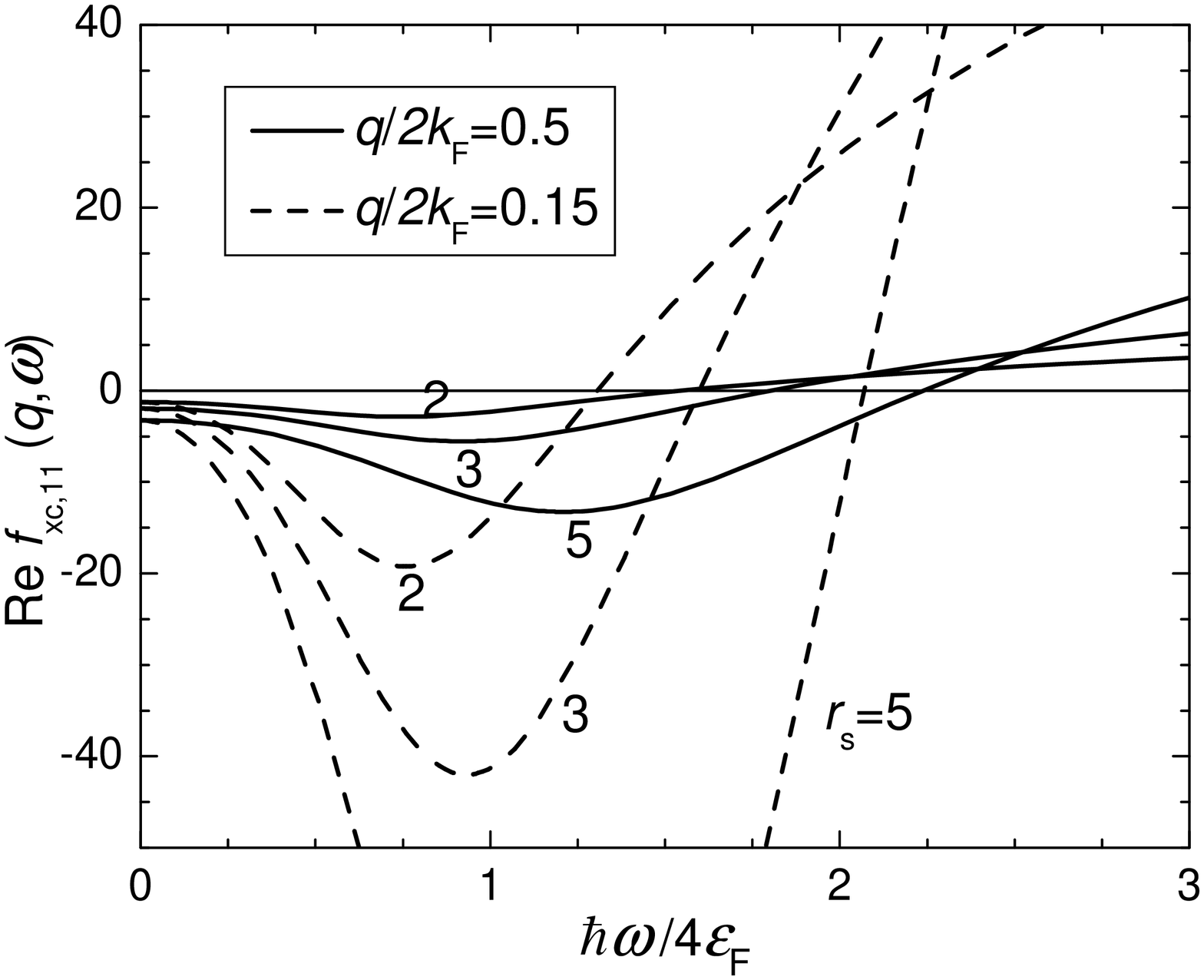}
\caption{Imaginary part (figure on the left) and real part (figure
on the
right) of the intra-layer dynamic xc kernel, $f_{\text{xc},11}(q,\protect%
\omega )$, of a bi-layer 2DES as a function of $\protect\omega $.
The kernels are shown for the total $r_{s}=2$, $3$, and $5$ and
for two different values of the
wave vector $q$ in the units of $g_{0}^{-1}$ where $g_{0}=m^{\ast }/\protect%
\pi \hbar ^{2} $ is the density of states for the noninteracting
2DES.} \label{fig1}
\end{figure*}

\begin{figure*}[h,t,b,p]
\centering
\includegraphics[width=6cm,angle=-90]{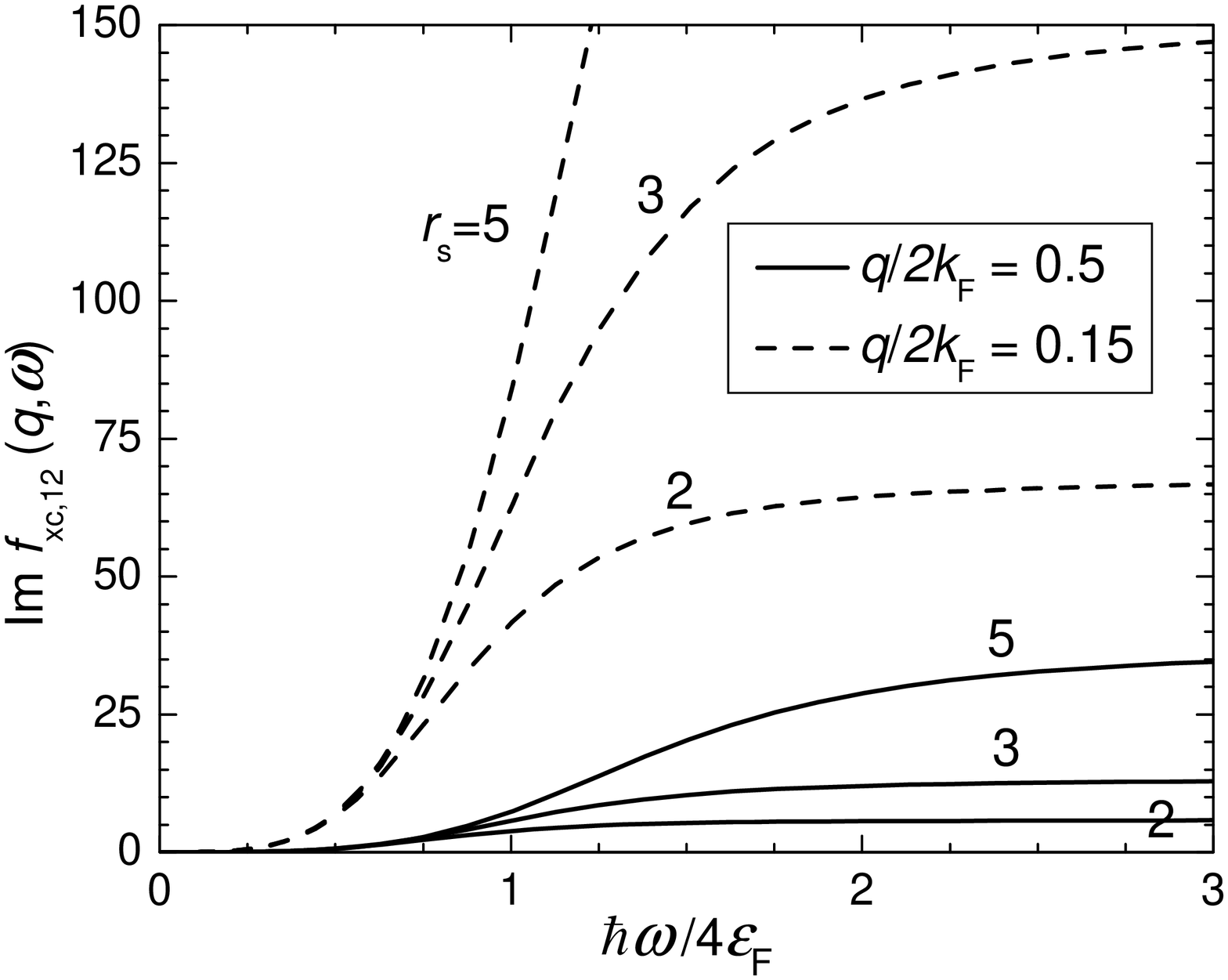} \hspace{1cm} %
\includegraphics[width=6cm,angle=-90]{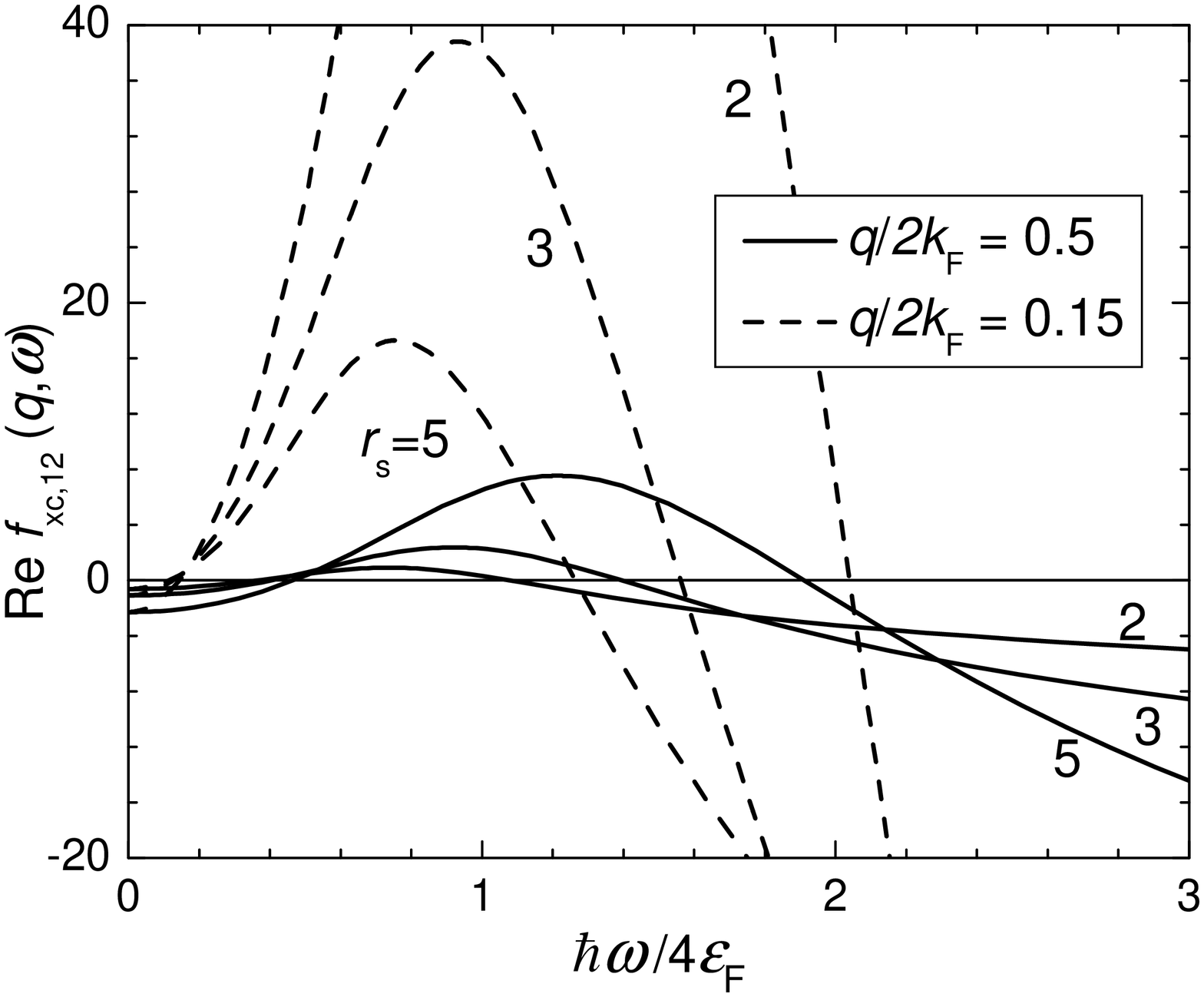}
\caption{Imaginary part (figure on the left) and real part (figure
on the right) of the inter-layer dynamic xc kernel,
$f_{\text{xc},12}(q, \omega )$, of a bi-layer 2DES as a function
of $\protect\omega$. The kernels are shown for the total
$r_{s}=2,~3$, and $5$ and for two different values of the wave
vector $q$ in the units of
$g_{0}^{-1}$ where $g_{0}=m^{\ast }/\protect%
\pi \hbar ^{2}$ is the density of states of the noninteracting
2DES.} \label{fig2}
\end{figure*}

Figs.~\ref{fig1} and \ref{fig2} show plots of the imaginary and
real parts of the intra- and inter-layer xc kernels,
$f_{\text{xc},11}(q,\omega )$ and $ f_{\text{xc},12}(q,\omega )$,
which we actually use in our calculations of the spectrum of
bi-layer plasmons and of the Coulomb drag mediated by these
plasmons. Notice that the kernels are calculated for the shown
values of $r_s=2,~3$ and $5$, which are obtained for the total
number of electrons, $n$, in a bi-layer. The corresponding
in-layer values of $r_s^{in-layer}$, calculated for the $n/2$
number of electrons in individual layers, are
$r_s^{in-layer}=2\sqrt{2},~3\sqrt{2}$ and $5\sqrt{2}$. At finite
frequencies, the term in Eq.~\ref{fxc}, related to $A(\omega )$,
dominates both in the short- and long-wavelength limits. The terms
related to $B^{\pm }(\omega )$ make essential contributions to
$f_{ \text{xc},ij}(q,\omega )$ only in the small $\omega $ region,
where $ A(\omega )$\ vanishes as $\omega^2$.

The frequency-dependence of the LFF in the particle-hole continuum
(PHC) region (finite wave vector, low frequency) is still largely
unknown. For this reason, in evaluating the particle-hole
contribution to the drag resistivity, we use the static limit of
the LFF. We take advantage of the analytical expressions recently
obtained for $G_{\text{xc}}^{-}(q)$ and $G_{\text{xc}}^{+}(q)$,
LFF \cite{davoudi}, which accurately reproduce the diffusion Monte
Carlo data \cite{senatore}, as well as the exact asymptotic
behaviors in the large and small $q$ limits. Notice that both the
dynamic xc kernels \cite{qv,qian} and the static LFF
\cite{davoudi}, which we have used in our calculations, are
evaluated at $T=0$ for an ideal 2DES of zero width. At present
there are no calculations of the LFFs of a two-dimensional system
that take into account the finite width of the well and/or the
temperature dependence. We expect, however, that  the temperature
dependence of the LFFs should play a relatively minor role in
comparison to the temperature dependence that we explicitly
include in the Fermi factors and the non-interacting polarization
functions.  Similarly, the effect of the finite width of the well
should largely be taken care of by our use of the form factors
$F_{ij}$ in Eq.~\ref{veff}.
\begin{figure*}[h,t,b,p]
\centering
\includegraphics[width=6cm,angle=-90]{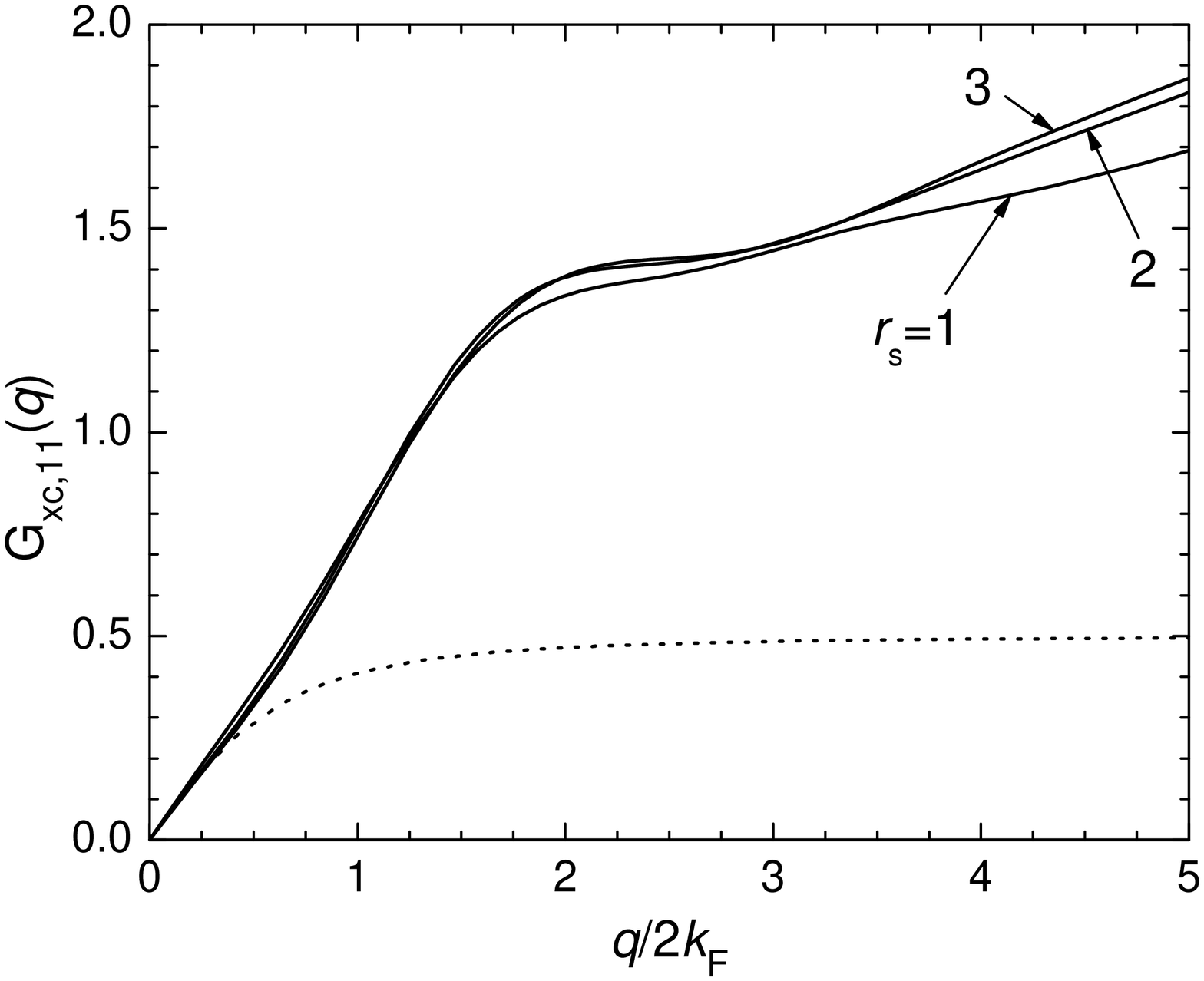} \hspace{1cm} %
\includegraphics[width=6cm,angle=-90]{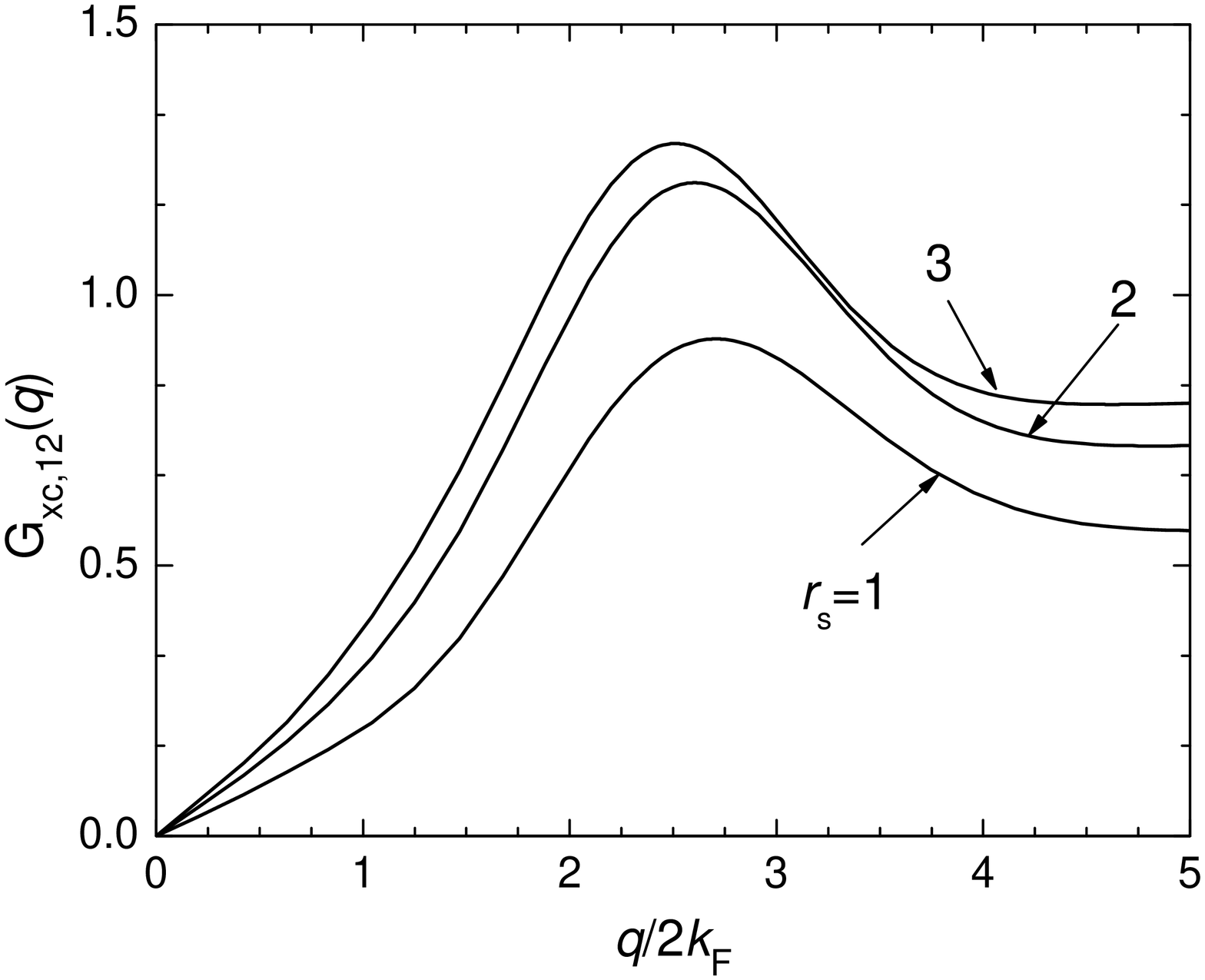}
\caption{Static many-body LFF of a double layer 2DES. The
intra-layer, $G_{\text{xc},11}$ (figure on the left), and the
inter-layer, $G_{\text{xc},12}$ (figure on the right), are plotted
as a function of wave vector $q$ for equal electron densities in
the two layers corresponding to the total $r_{s}=1$, $2$, and $3$.
The dotted curve shows the LFF $G_H (x)$ in the Hubbard
approximation.} \label{fig3}
\end{figure*}
In Fig.~\ref{fig3} we plot the intra- and inter-layer LFF
$G_{{\text{xc}},ij}(q)$ which we actually use in our calculations
of the drag resistivity in the PHC region. For all the values of
$r_{s}$ shown in the figure, the intra-layer LFF
$G_{{\text{xc}},11}(q)$ become significantly larger than the LFF
in the Hubbard approximation, $G_{\text{H}}(x)=x/\sqrt{1+4x^{2}}$
when $x$ is close to unity. This makes the effective intra-layer
interaction weak in the region near $x=1$ and results in a weak
in-plane screening effect. This effect becomes especially
important in the regime of low density bi-layers where the main
contribution to drag comes from the large-angle scattering
processes with $x\approx 1$. In Fig.~\ref{fig3} we also see that
although $G_{\text{xc},12}(q)$ is smaller than
$G_{\text{xc},11}(q)$, the two LFF are of the same order of
magnitude.

\section{Bi-layer plasmon dispersions: Effect of dynamic xc}

Below we consider the symmetric bi-layer 2DES ($n_1=n_2$) with
inter-layer spacing $\Lambda =28$ nm and quantum wells width
$d=18$ nm, which correspond to the experimental situation of
Ref.~\onlinecite{kellogg}.

In this section we calculate the spectrum of bi-layer plasmons for
the total $r_{s}=2$ and $3$ ($r_s^{in-layer}=2\sqrt{2}$ and
$3\sqrt{2}$) both within the RPA and beyond it taking into account
the dynamic xc effects through the intra- and inter-layer xc
kernels. In the symmetric systems the bi-layer screening function
is represented as
\begin{equation}
\varepsilon_{\text{bi}}(q,\omega )=\varepsilon _{+}(q,\omega
)\varepsilon_{-}(q,\omega )~,
\end{equation}
where
\begin{equation}
\varepsilon _{\pm }(q,\omega )=1+V_{\pm }(q,\omega )\Pi ^{0}(q,\omega )~,
\end{equation}
with $V_{\pm }(q,\omega )=V_{\text{eff},11}(q,\omega )\pm V_{\text{eff}%
,12}(q,\omega )$. The spectrum of the collective excitations is
obtained from the zeros of the real part of the bi-layer screening
function so the dispersion is obtained from the solution of the
equation
\begin{equation}  \label{dispersionequation}
\text{Re}~\varepsilon_{\pm }(q,\omega _{\text{op,ac}}(q))=0~.
\end{equation}
The imaginary part of the screening function describes the damping of the
collective modes. In this approach the imaginary parts of the dynamic xc
kernels are finite at finite frequencies and we obtain that even at zero
temperature the collective excitations are Landau-damped outside the $T=0$
PHC region. Also, we use the finite temperature forms for the polarization
function $\Pi ^{0}(q,\omega )$ which itself has a twofold effect on the
plasmon spectrum. The finite temperature contribution to the imaginary part
of $\Pi ^{0}(q,\omega ) $ outside the $T=0$ PHC region enhances the damping
of the collective modes, while temperature-induced changes in the real part
of $\Pi^{0}(q,\omega )$ modify the domain of existence of the collective
excitations.

\begin{figure*}[h,t,b,p] \centering
\includegraphics[width=6cm,angle=-90]{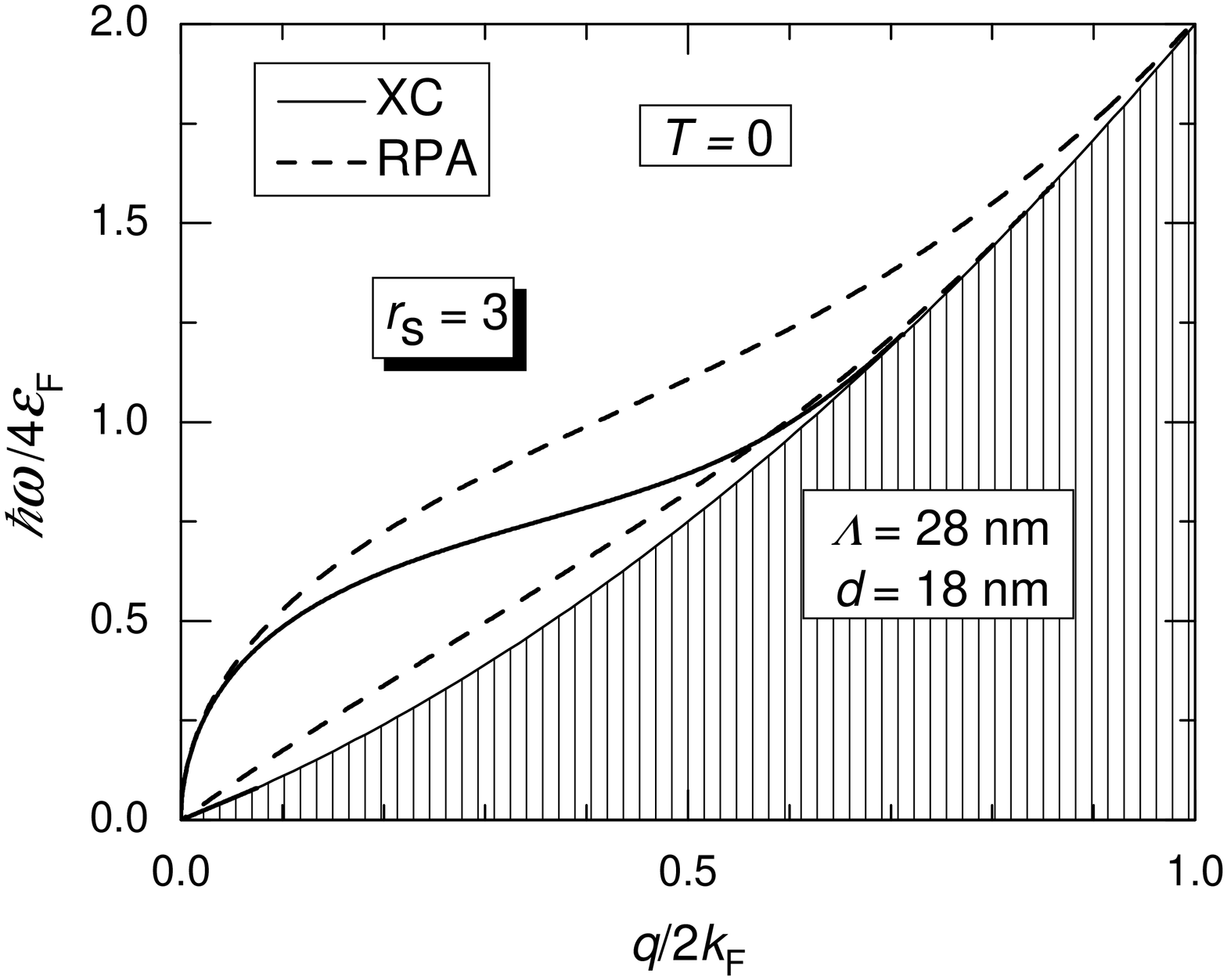} \hspace{1cm}%
\includegraphics[width=6cm,angle=-90]{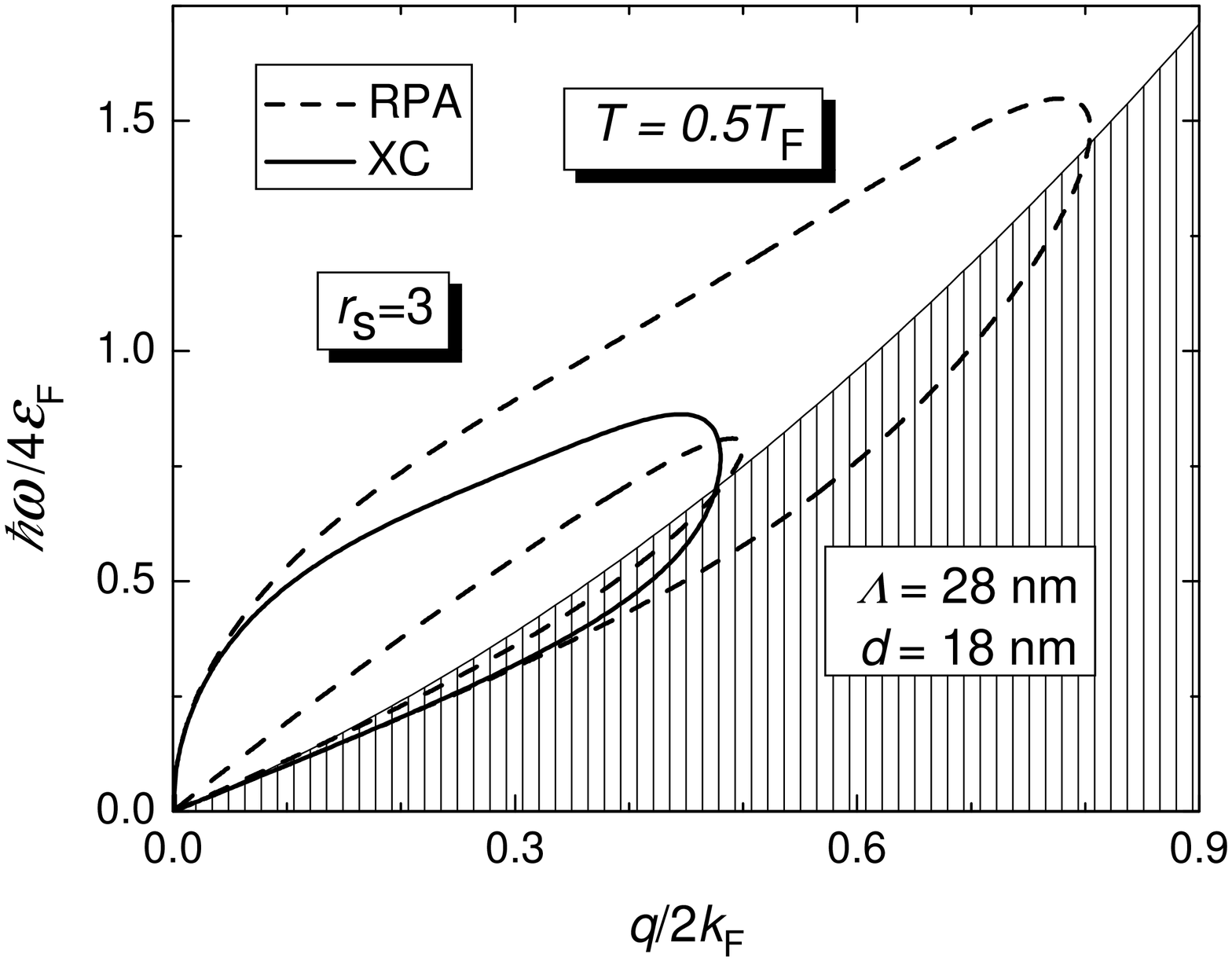}
\caption{Bi-layer plasmon spectra at zero temperature (figure on
the left) and at finite temperature, $T=0.5 T_F$ (figure on the
right) for the total $r_{s}=3$. The hatched area shows the PHC
region at $T=0$. The dispersions curves are calculated within the
RPA (dashed curves) and beyond the RPA, taking into account the
full intra- and inter-layer dynamic xc corrections (solid
curves).} \label{fig4}
\end{figure*}

In Fig.~\ref{fig4} we plot the plasmon spectra at zero and finite
temperatures for the total $r_{s}=3$, based on the above
Eq.~(\ref{dispersionequation}). The spectra are calculated within
the RPA and beyond it taking into account the full intra- and
inter-layer xc kernels. The upper-lying  ``optical" branches in
both approximations exhibit the expected square-root $\sim
\sqrt{q}$ dispersion in the long wavelength limit, while the lower
lying "acoustical" branches have the linear $\sim q$ dispersion.
The inclusion of the dynamic xc kernels forces both the optical
and acoustical branches to enter the PHC at substantially smaller
values of $q$. At finite $T$ the dispersions of collective
excitations show the same behavior in the long wavelength limit
while at the upper edge of the wave vector interval in which the
collective excitations exist, the group velocity of each type of
plasmon, independently of the approximation used, becomes
infinite.

\begin{figure*}[h,t,b,p]
\centering
\includegraphics[width=6cm,angle=-90]{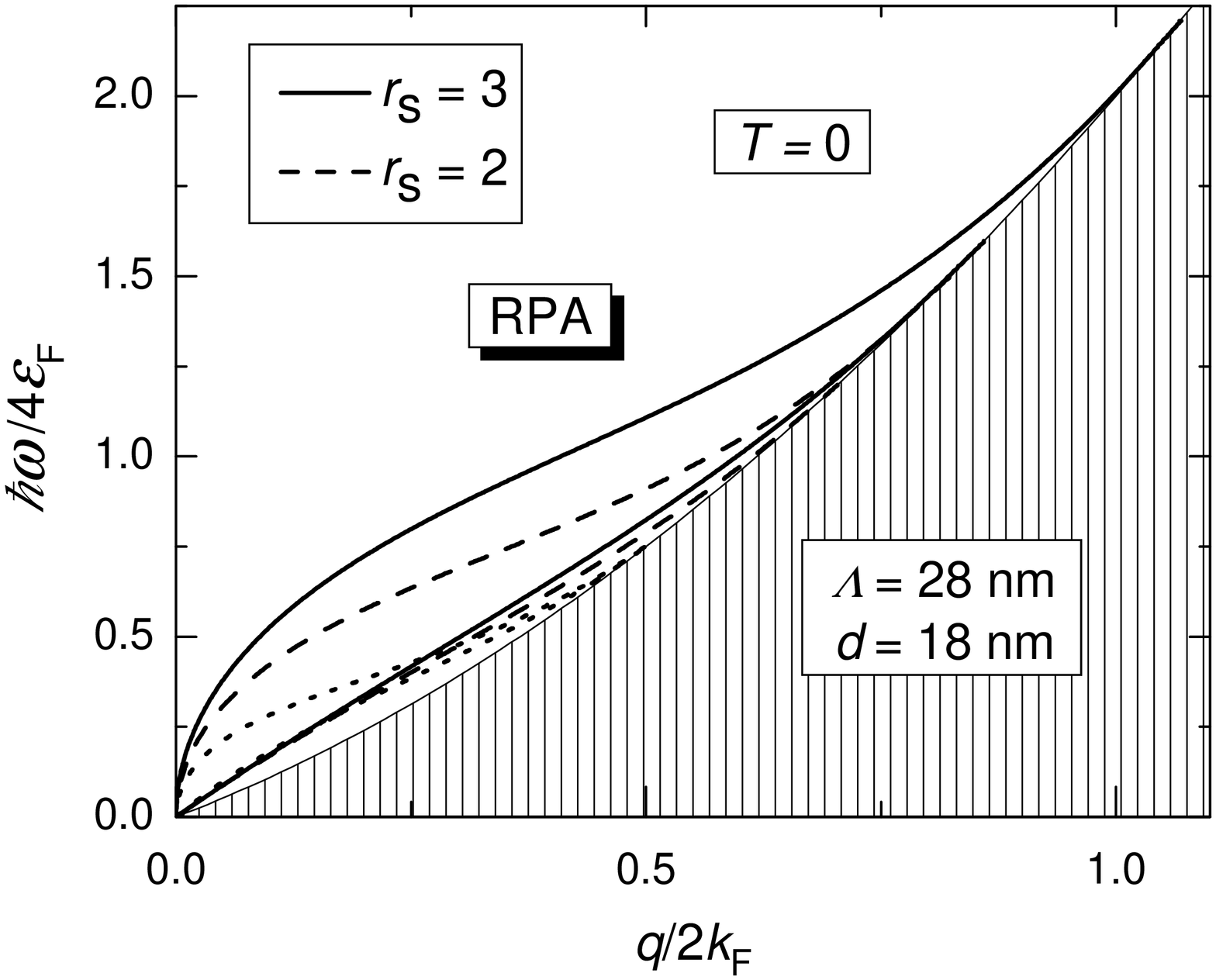} \hspace{1cm} %
\includegraphics[width=6cm,angle=-90]{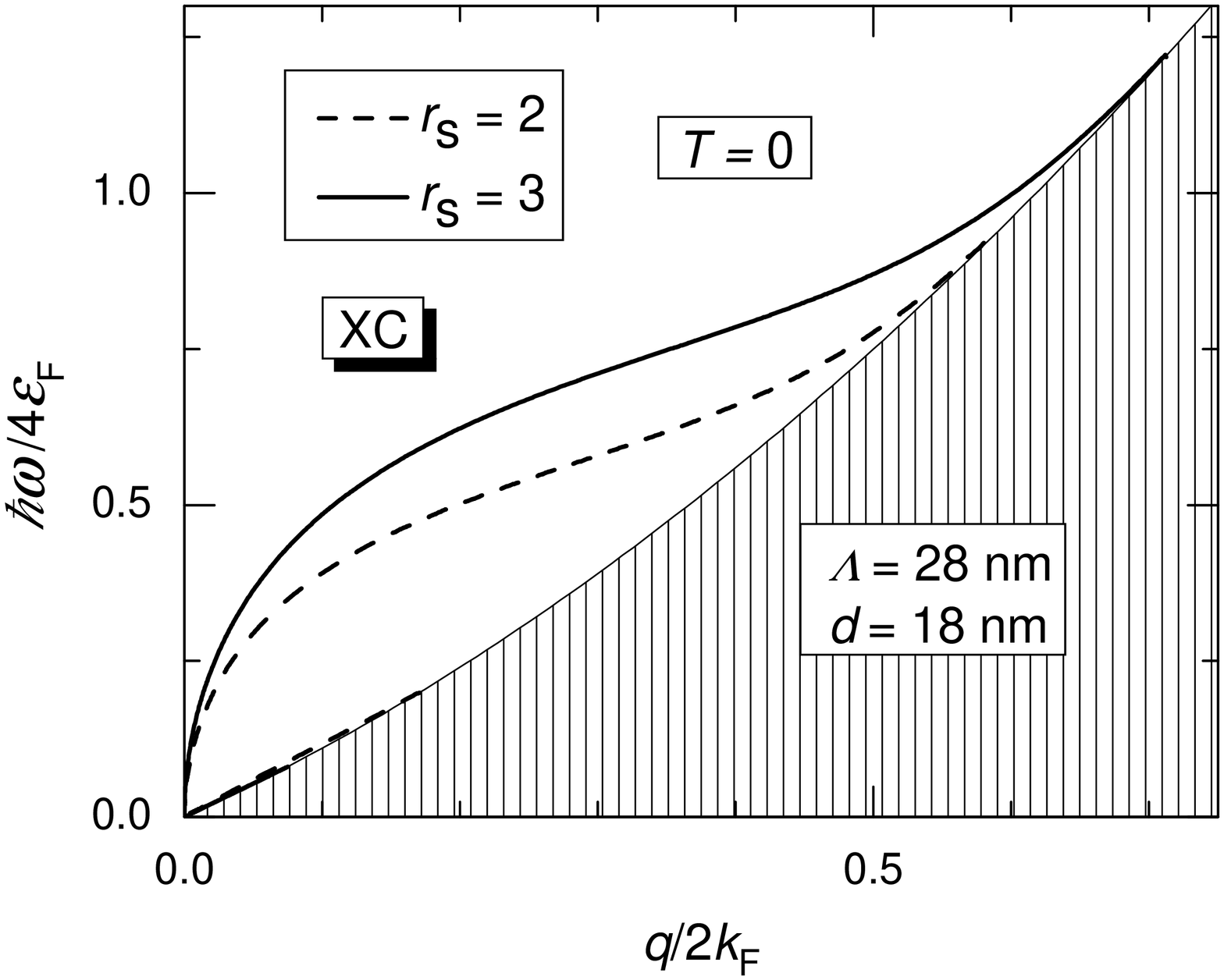}
\caption{Bi-layer plasmon spectra at zero temperature. The hatched
area shows the PHC region at $T=0$. The dispersion curves are
calculated within the RPA (figure on the left) and beyond the RPA
by taking into account the effects of intra- and inter-layer
dynamic xc (figure on the right). The solid and dashed curves
correspond to the values of the total $r_{s}=2$ and $3$,
respectively. The dotted curves reproduce the plasmon dispersions
from Ref.~\onlinecite{flensberg}, calculated within the RPA for
the high density, $n=1.5\cdot 10^{11}$ cm$^{-2}$, bi-layer samples
with the inter-layer separation $k_F \Lambda=4$. } \label{fig5}
\end{figure*}

Fig.~\ref{fig5} shows that, as the electron density decreases,
both the optical and the acoustical plasmon dispersions,
calculated within the RPA, move away from the boundary of the PHC.
We find, however, that the many-body effects beyond the RPA result
in a new tendency, namely, the optical and the acoustical modes
repel each other when $r_{s}$ increases so that the acoustical
branch becomes closer to the boundary of the PHC while the optical
branch is still pushed away from it. As we see below all these
changes, induced by the dynamic many-body effects, are directly
reflected in the temperature dependence of the plasmon-mediated
Coulomb drag.

\section{Drag resistivity within the RPA}

\begin{figure}[h,t,b,p]
\centering\includegraphics[width=6.5cm,angle=-90]{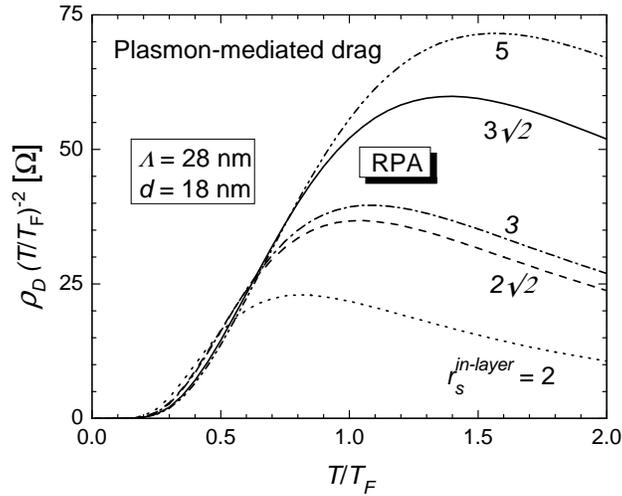}
\caption{Plasmon-mediated drag resistivity as a function of scaled
temperature within the RPA. The curves  from bottom to top
correspond to the in-layer
$r_{s}^{in-layer}=\sqrt{2},~2,~2\sqrt{2},~3,~3\sqrt{2}$, and $5$.}
\label{fig6}
\end{figure}

First we present the RPA based calculations of the drag due to the
exchange of plasmons in low-density bi-layers. In Fig.~\ref{fig6}
the transresistivity (scaled by the factor $(T/T_F)^2$) is plotted
as a function of temperature for five different in-layer densities
corresponding to $r_{s}^{in-layer}=2,~2\sqrt{2},~3,~3\sqrt{2}$,
and $5$. Although the magnitude of the drag resistivity increases
with $r_{s}$, it is seen that within the RPA the qualitative
behavior of the plasmon-mediated drag does not undergo strong
changes when the electron density decreases, and it resembles the
behavior at high densities \cite{flensberg}. The behavior of the
transresistivity versus temperature exhibits an upturn at low
temperatures near $0.2T_{F}$ and a plasmon enhancement peak at
higher temperatures. At low temperature the transresistivity
decreases slightly with decreasing density. In contrast to this,
at high temperatures the height of the transresistivity peak
increases approximately linearly with $r_{s}^{in-layer}$ in the
range $2<r_{s}^{in-layer}<5$. As seen in Fig.~\ref{fig6} both the
upturn temperature of the plasmon-mediated drag and the position
of the plasmon enhancement peak increase with $r_{s}^{in-layer}$.
This is consistent with the above discussion of the plasmon
spectra, since in the RPA the plasmon dispersion curves for both
modes move away from the PHC when the density decreases. Notice,
however, that the absolute value of the upturn temperature and the
position of the plasmon-enhancement peak (not scaled by $T_{F}$)
show opposite tendency and decrease with $r_{s}^{in-layer}$. This
is due to the fact that the Fermi temperature, $T_{F}$, decreases
quadratically with $r_{s}^{in-layer}$.

\begin{figure}[h,t,b,p]
\centering\includegraphics[width=6.5cm,angle=-90]{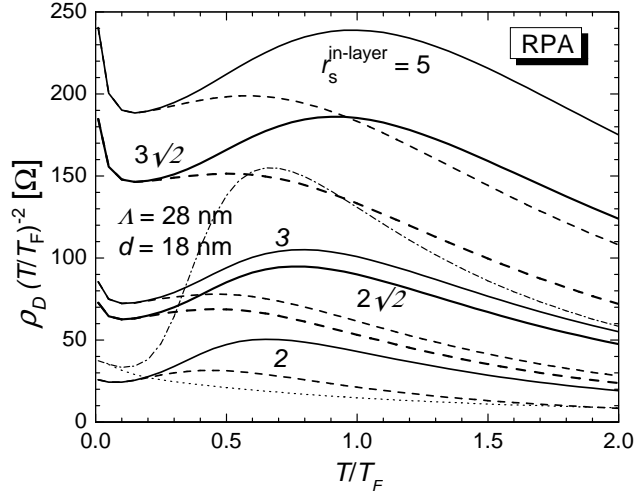}
\caption{Coulomb drag resistivity as a function of scaled
temperature within the RPA for the in-layer
$r_{s}=2,~2\sqrt{2},~3,~3\sqrt{2}$, and $5$ shown, respectively,
from bottom to top. The solid curves correspond to the total drag,
mediated both by plasmon and by particle-hole excitations, while
the dashed curves represent only the particle-hole contribution.
The dashed-dotted and dotted curves are for the situation
corresponding to the first drag measurements of
Ref.~\onlinecite{first}: they are rescaled by a factor of $20$,
and represent, respectively, the total drag and the drag due to
exchange of particle-hole excitations only. The latter is
calculated within the static screening approximation.}
\label{fig7}
\end{figure}

In Fig.~\ref{fig7} we plot the total transresistivity, due to
exchange of both plasmons and particle-hole excitations against
the part of the transresistivity that is due only to exchange of
particle-hole excitations within the RPA. One can see that for
$r_{s}$= $2,~2\sqrt{2},~3,~3\sqrt{2}$, and $5$ when the density
decreases, the contribution to the drag, made by particle-hole
excitations, increases faster than the plasmon contribution to
drag. However, the plasmon contribution remains essential in all
cases and results in the plasmon enhancement peaks, clearly seen
in the behavior of the scaled transresistivity versus temperature.

The dotted and dashed-dotted curves in Fig.~\ref{fig7} reproduce,
respectively, the calculations by Jauho and Smith of the
particle-hole contribution to the transresistivity based on the
static screening approximation \cite{jauho} and by Flensberg and
Hu of the plasmon-enhanced drag \cite{flensberg} in the dynamic
RPA. These curves correspond to the situation of the first drag
experiment \cite{first} on high density samples with $n=1.5\times
10^{11}$ cm$^{-2}$ and with a large center-to-center inter-layer
spacing, $\Lambda =37.5$ nm. Notice that the plasmon enhancement
peak is most pronounced in Fig.~\ref{fig7}. However, this strong
effect is mainly caused by the large inter-layer separation which
exceeds significantly the effective spacing of the bi-layers in
the experiment in Ref.~\onlinecite{kellogg} for which the solid
lines are calculated. As far as the plasmon-mediated drag depends
more weakly on inter-layer spacing than the particle-hole
contribution to drag \cite{flensberg}, the plasmon enhancement
peak is more pronounced for the dotted curves.

\section{Drag resistivity beyond the RPA}

\subsubsection{The effect of dynamic xc on plasmon-mediated drag}

\begin{figure}[h,t,b,p]
\centering\includegraphics[width=4.2cm,angle=-90]{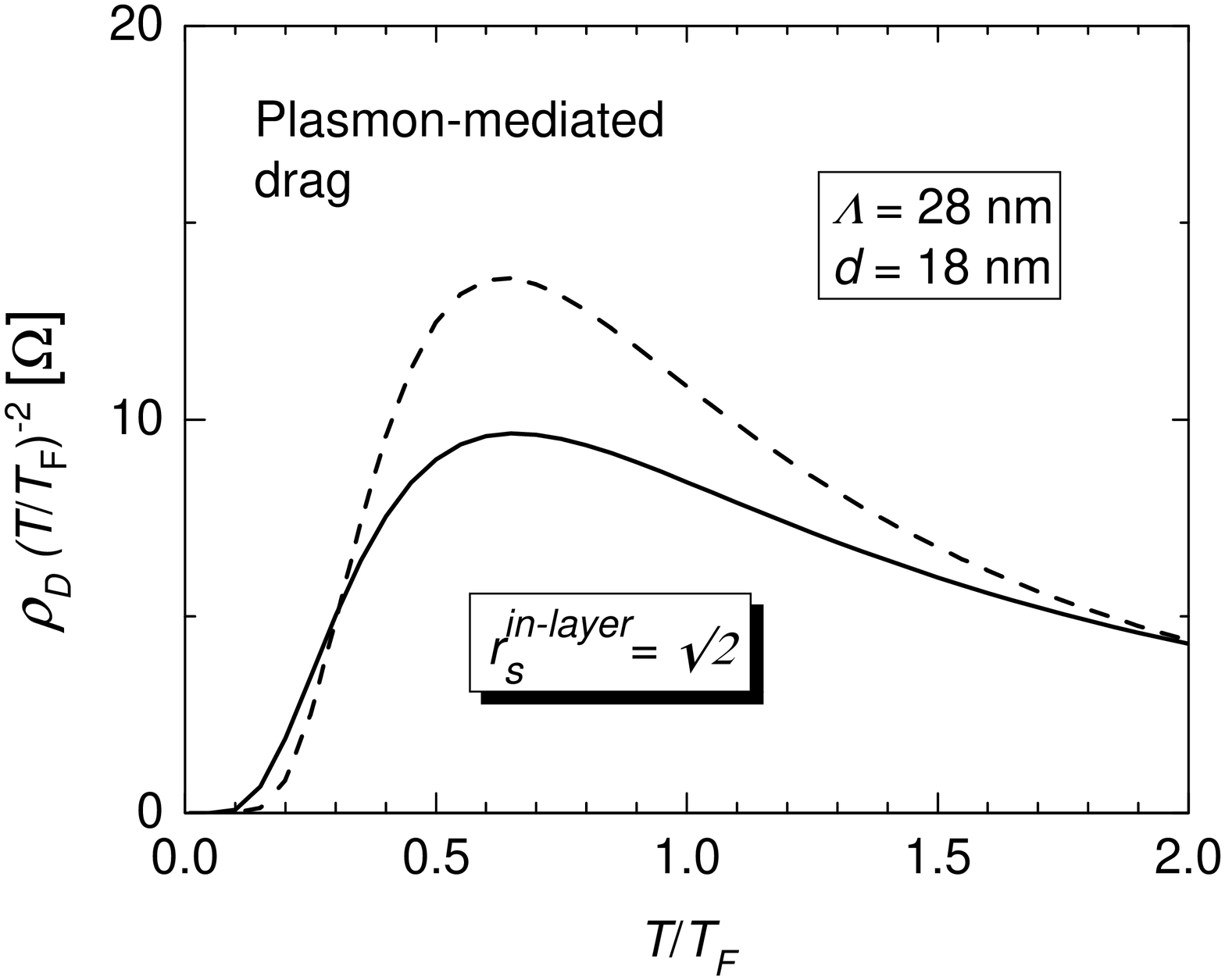}
\centering\includegraphics[width=4.2cm,angle=-90]{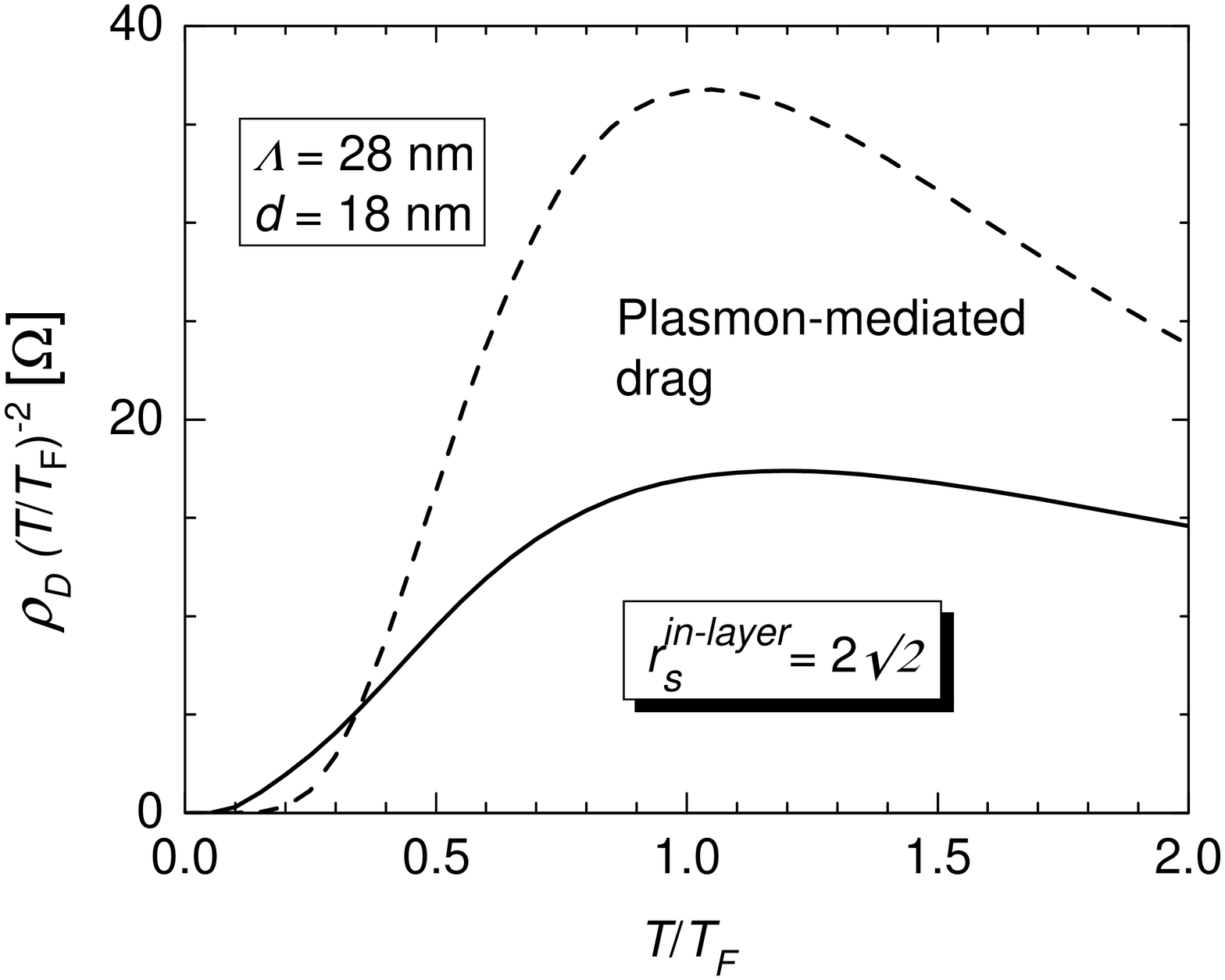}
\centering\includegraphics[width=4.2cm,angle=-90]{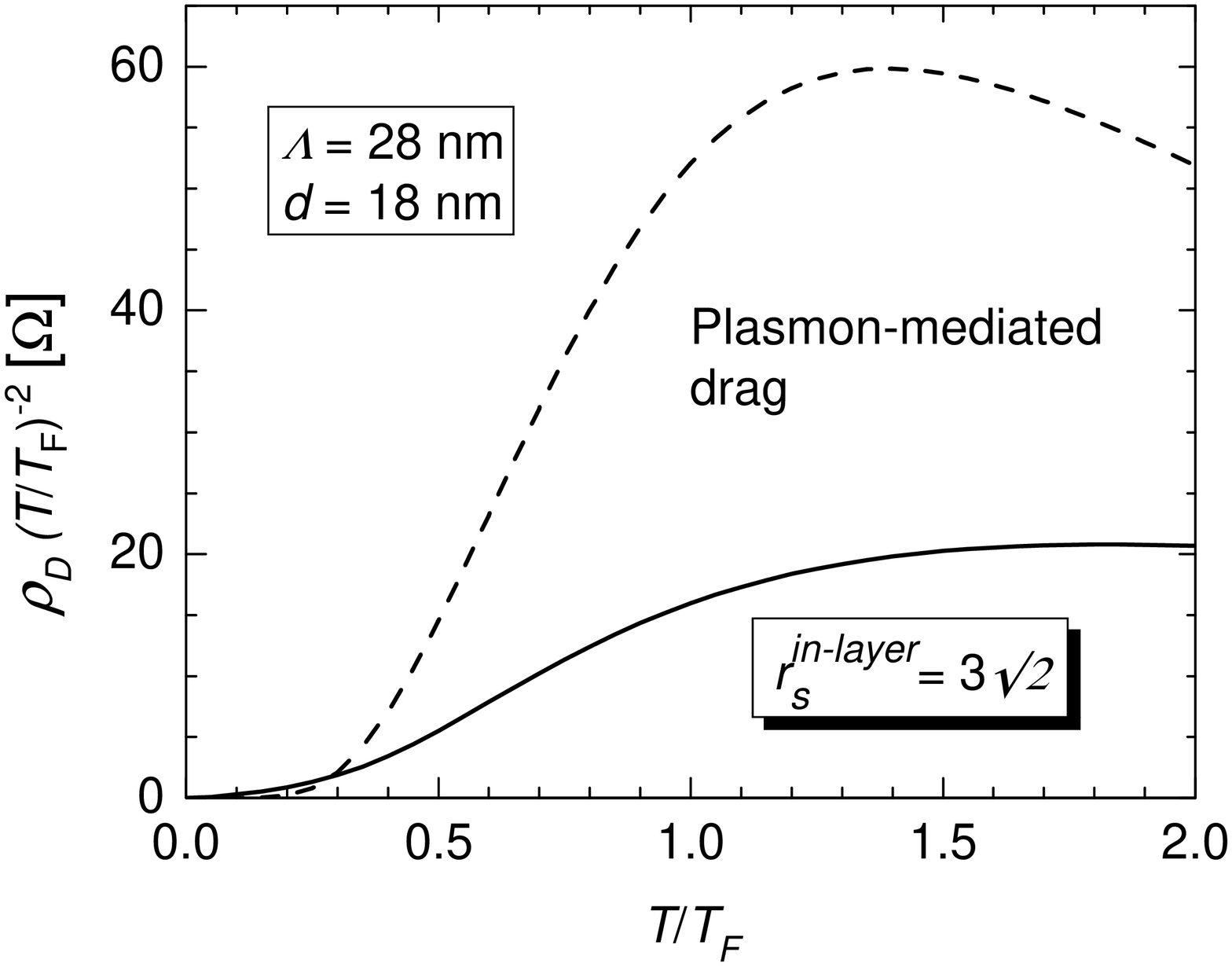}
\caption{Dynamic xc corrections to the plasmon-mediated drag for
the in-layer $r_{s}^{in-layer}=\sqrt{2},~2\sqrt{2}$, and
$3\sqrt{2}$. The scaled transresistivity vs temperature is shown
within the RPA (dashed curves) and beyond the RPA taking into
account the intra- and inter-layer xc effects (solid curves).}
\label{fig8}
\end{figure}

In this subsection we include the intra- and inter-layer dynamic
xc within the scheme (\ref{dyson})-(\ref{veff}) and study their
effect on the plasmon-mediated drag by comparing the obtained
results with those of RPA-based calculations. In Fig.~\ref{fig8}
the scaled transresistivity as a function of temperature is shown
for three equal density bi-layer systems with the total
$r_{s}=1,~2$ and $3$ corresponding to the in-layer
$r_{s}^{in-layer}=\sqrt{2},~2\sqrt{2}$ and $3\sqrt{2}$. It is seen
that the upturn temperature in the scaled transresistivity
essentially decreases when the xc effects are included. Despite
slight differences, in both cases of $r_{s}^{in-layer}=\sqrt{2}$
and $2\sqrt{2}$ the plasmons begin to contribute heavily to the
inter-layer e-e interaction at temperatures less than $0.1T_{F}$.
At $r_{s}^{in-layer}=3\sqrt{2}$ the effect becomes even stronger.
Whereas, according to the RPA calculation, the upturn of the
plasmon-mediated drag should be about $0.2T_{F}$. This difference
is explained by the changes in the spectrum of bilayer plasmons
induced by the dynamic xc effects and discussed in the previous
section. At low temperature the plasmon-mediated drag is mainly
determined by the acoustical plasmons, whose frequency is reduced
by the dynamic xc: this makes the acoustic plasmons easier to
excite thermally, and causes their contribution to the Coulomb
drag to be larger than in RPA.

When temperature increases, the average energy of plasmons that
mediate the drag increases. Hence, the plasmon damping, caused by
dynamic xc effects, becomes stronger with a consequent reduction
in the drag transresistivity. Thus, at temperatures around
$0.3T_F$ the plasmon-mediated drag obtained within the RPA exceeds
the drag which takes into account the xc effects.

At still higher temperatures the plasmon-mediated drag shows a
broad peak both within the RPA and beyond it. The position of this
peak is determined mainly by the position of the optical plasmons
insofar as at such high temperatures the acoustical plasmons are
heavily damped and merged into the PHC. As seen from
Figs.~\ref{fig8} the peak position is at higher temperature at the
lower density, and it changes slightly depending on the chosen
approximation. These observations are again consistent with the
behavior of the plasmon spectra in Figs.~\ref{fig4} and
\ref{fig5}. The main contribution to the plasmon-mediated drag is
made by the plasmons with energies $\hbar \omega \sim T$ while the
scaled transresistivity peaks occur at temperatures $T\lesssim
T_{F}$. Therefore, as seen in Figs.~\ref{fig4} and \ref{fig5}, at
such small values of $\hbar \omega /4\varepsilon _{F}\lesssim
0.25$, the positions of the optical plasmons, and hence the
positions of the peak in the plasmon contribution to drag in
Figs.~\ref{fig8}, remain approximately unaffected. Notice also
that at even higher temperatures the differences in the magnitude
of plasmon-mediated drag in the different approximations
diminishes with increasing $T$.

\subsubsection{The combined effect of dynamic and static xc on
the total drag}

\begin{figure}[h,t,b,p]
\centering\includegraphics[width=6cm,angle=-90]{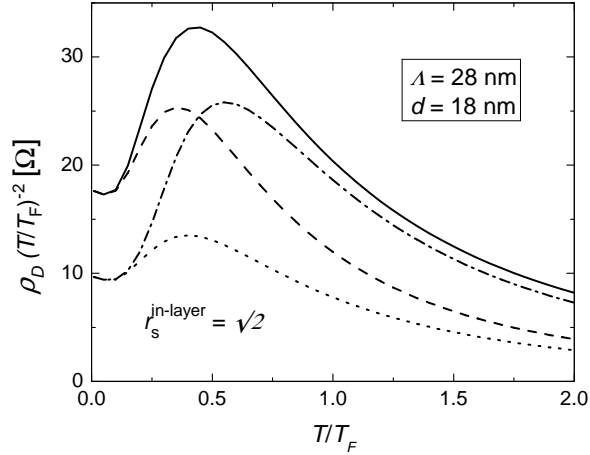}
\caption{The drag resistivity vs temperature for the in-layer
$r_{s}^{in-layer}=\sqrt{2}$. The solid and dashed curves represent
the total transresistivity and the separate particle-hole
contribution to the drag resistivity with the full intra- and
inter-layer xc corrections. The dash-dotted and dotted curves are,
respectively, the total transresistivity and the particle-hole
contribution to the drag calculated within the RPA.} \label{fig10}
\end{figure}

In this subsection we investigate the combined effect of the
dynamic and static many-body xc on the total drag, mediated both
by plasmon and by particle-hole excitations. In Fig.~\ref{fig10}
the drag resistivity is plotted vs temperature within and beyond
the RPA for the total $r_{s}=1$ ($r_{s}^{in-layer}=\sqrt{2}$). It
is seen that in both approximations the particle-hole contribution
to the scaled drag resistivity first shows a slight dip followed
by a peak at higher temperatures. In the RPA the plasmon
contribution to drag enhances this peak and shifts it to even
higher temperatures so that the total transresistivity shows a
pronounced peak approximately at the position where the plasmon
contribution has a peak. As seen from Fig.~\ref{fig10} the
introduction of the static exchange-controlled LFF increases the
peak height of the particle-hole contribution to the drag and
shifts the peak towards lower temperatures. On the other hand, the
plasmon-mediated drag is moderately suppressed by the dynamic xc
corrections as discussed in the previous subsection (cf.
Fig.~\ref{fig8}). Thus, the resulting peak in the graph of the
total transresistivity vs temperature becomes smaller at
relatively low temperatures, while at high temperatures the drag
rate shows a monotonic decrease in $T$.

\begin{figure}[h,t,b,p]
\centering
\includegraphics[width=6cm,angle=-90]{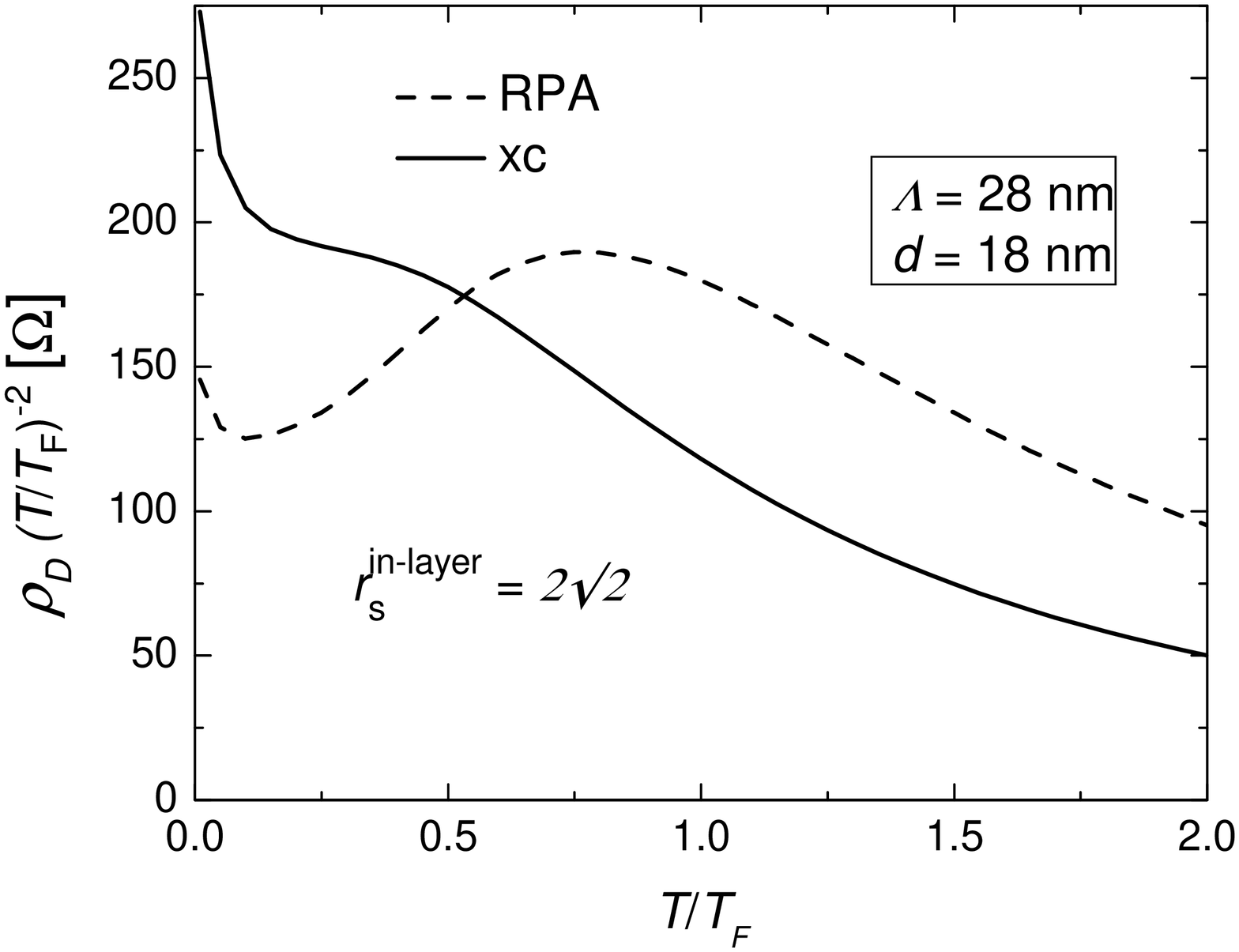}
\includegraphics[width=6cm,angle=-90]{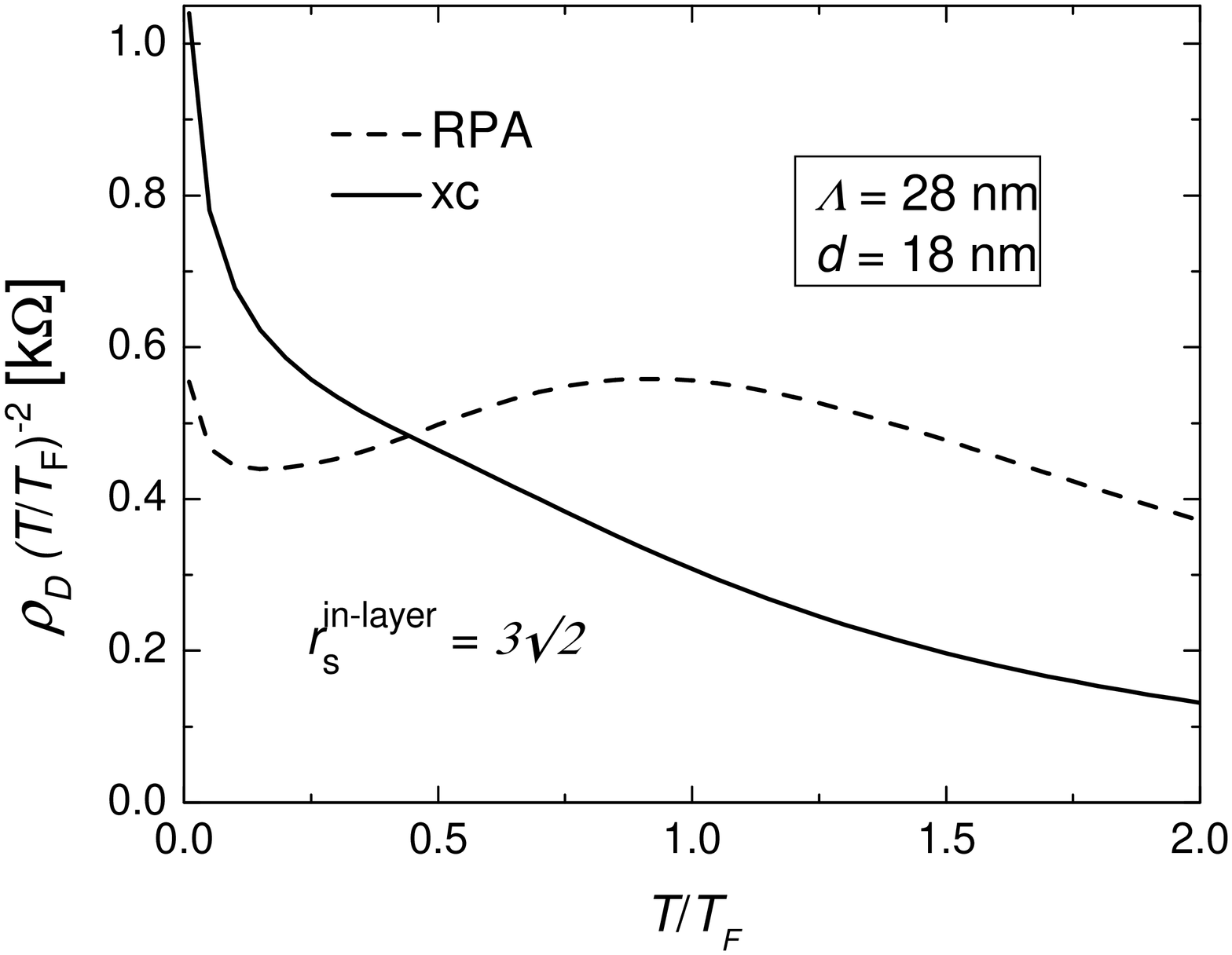}
\caption{The full xc corrections to the total transresistivity for
the in-layer $r_{s}^{in-layer}=2\sqrt{2}$ and $3\sqrt{2}$. The
dashed and solid curves represent the scaled transresistivity vs
temperature, calculated, respectively, within and beyond the RPA.
The RPA data are multiplied by a factor of $3$ and $6$,
respectively for $r_{s}^{in-layer}=2\sqrt{2}$ and
$r_{s}^{in-layer}=3\sqrt{2}$.} \label{fig11}
\end{figure}

The described quantitative and qualitative differences in the
behavior of the scaled total transresistivity within and beyond
the RPA becomes more pronounced at lower densities. As seen from
Fig.~\ref{fig11} for the in-layer $r_{s}^{in-layer}=2\sqrt{2}$ and
$3\sqrt{2}$ the total transresistivity beyond the RPA as a
function of temperature shows no peak at all and this is in stark
contrast to the peaked behavior of the transresistivity within the
RPA. The disappearance of the large high-temperature plasmon peak
results from the strong increase of the drag resistivity at low
temperatures: we ascribe this to the fact that the contribution to
drag, made by large-angle inter-layer scattering processes,
becomes dominant when the many-body xc corrections are included.

\begin{figure}[h,t,b,p]
\centering\includegraphics[width=6cm,angle=-90]{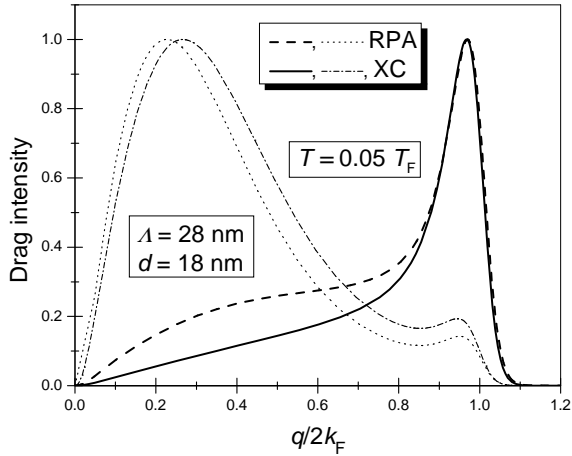}
\caption{Drag intensity $I(x)$ as a function of transferred
momentum $ x=q/2k_F$ for a low density $n=3.8\cdot10^{10}$
cm$^{-2}$ bi-layer. For comparison the drag intensity is plotted
also for a high density $n=1.5\cdot10^{11}$ cm$^{-2}$ bi-layer.
The dashed (dotted) and solid (dash-dotted) curves represent
respectively the results of calculations within and beyond the RPA
at $T=0.05 T_F$ for $n=3.8\cdot10^{10}$ cm$^{-2}$
($n=1.5\cdot10^{11}$ cm$^{-2}$). The respective values of the drag
resistivity are given by the area under these curves. The curves
are normalized to their particular peak values,
$\rho^{RPA}_{_{p}}=0.57~(0.06)$ and
$\rho^{xc}_{_{p}}=2.71~(0.01)~\Omega$ for $n=3.8\cdot10^{10}$
cm$^{-2}$ ($n=1.5\cdot10^{11}$ cm$^{-2}$).} \label{fig12}
\end{figure}

As shown recently in Ref.~\onlinecite{kellogg}, the large-angle
scattering processes with $q\simeq 2k_{F}$ play an important role
in the Coulomb drag in low density bi-layers. In such samples with
sufficiently small inter-layer separations, $k_{F}\Lambda $ is so
small that the usual exponential cutoff, $exp(-q\Lambda )$, of the
inter-layer e-e interaction is no longer effective in suppressing
the large-angle scattering processes. Thus, according to previous
predictions\cite{wilkins}, the large-angle inter-layer scattering
events, due to the divergence of the scattering phase-space near
$q\simeq 2k_{F}$, lead to a $T^{2}lnT$ behavior of the drag
resistivity at low temperatures, in lieu of the usual $T^{2}$
temperature dependence, which is driven by small-angle scattering
processes. To distinguish clearly the contributions to the drag
made by the small- and large-angle inter-layer scattering
processes, in Fig.~\ref{fig12} we have plotted the dimensionless
drag intensity, $I(x)$, within and beyond the RPA versus the
average transferred momentum, $x=q/2k_F$, at low temperature
$T=0.05T_{F}$. We have normalized the drag intensity to its
corresponding peak values, $\rho_{_{p}}$, so that the
transresistivity $\rho_{_{D}}=\rho_{_{p}}\int I(x) dx$. It is
evident that the xc effects at the low density reduce essentially
the component from the small-angle scattering processes so that
the large-angle scattering contribution dominates the drag. Thus,
the drag rate is strongly enhanced at low temperatures owing to
the logarithmic corrections provided by the large-angle scattering
events. By contrast, in the RPA the contribution of the
small-angle scattering processes remains dominant down to the
lowest temperatures (see Fig.~\ref{fig12}).

\begin{figure}[h,t,b,p]
\centering
\includegraphics[width=6cm,angle=-90]{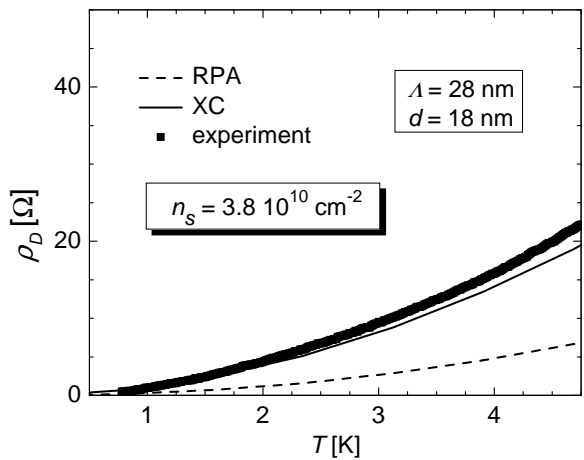}
\caption{The drag resistivity versus temperature for $n=3.8\cdot
10^{10}$ cm$^{-2}$. The solid curve represents calculations when
the intra- and inter-layer many-body corrections are included, the
dashed curve is the RPA results. The symbols are the experimental
findings from Ref.~ \onlinecite{kellogg} .} \label{fig13}
\end{figure}

Lack of experimental measurements of the Coulomb drag at high
temperatures does not allow for the time being an experimental
verification of our predictions on the position and strength of
the plasmon peak in low density bi-layers. Notice however that at
low temperatures, as shown in Fig.~\ref{fig13}, our numerical
results are in good agreement with the experimental findings by
Kellogg \textit{et al.}~\cite{kellogg} for $ n=3.8\times
10^{10}$cm$^{-2}$. In this figure we have shown the drag
resistivity due to exchange of particle-hole excitations.
Inclusion of the dynamical part will only slightly increase the
transresistivity since at such low temperatures plasmons only
start to contribute. As seen in Fig.~\ref{fig13} the RPA
calculations near $T=5$ K underestimates the experimental results
more than three time. Meanwhile after inclusion of the xc
corrections, the drag resistivity is about $88\%$ of the
experimental result. This is very good agreement against the
background of the usual discrepancy between theory and experiment
in determining the magnitude of the drag.

\begin{figure}[h,t,b,p]
\centering\includegraphics[width=6cm,angle=-90]{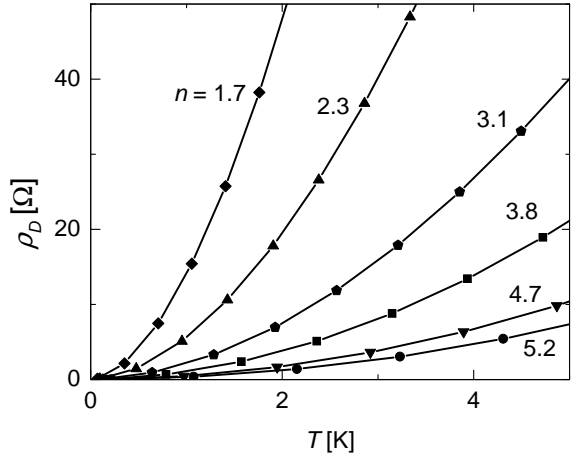}
\caption{The drag resistivity vs temperature for six different
densities corresponding to Fig.~1 from Ref.~\onlinecite{kellogg}.
The densities are in units of $10^{10}$ cm$^{-2}$.} \label{exp}
\end{figure}

\begin{figure}[h,t,b,p]
\centering\includegraphics[width=6cm,angle=-90]{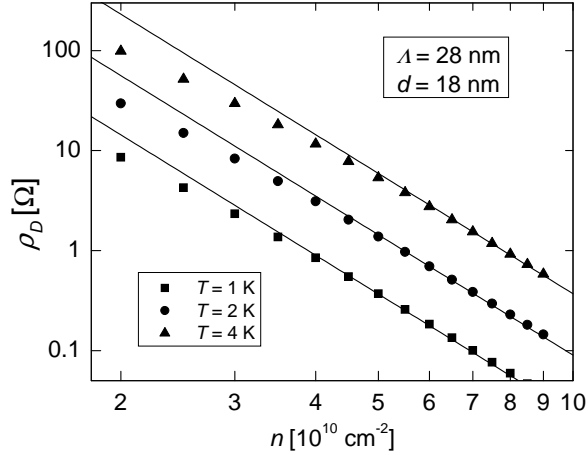}
\caption{Log-log plot of the transresistivity vs density for three
different temperatures: $T=1,2$, and $4$ K. This figure
corresponds to Fig.~2 from Ref.~\onlinecite{kellogg}. The solid
lines are proportional to $n^{-4}$.} \label{lglg}
\end{figure}

In Fig.~\ref{exp} we plot the drag resistivity versus temperature
for six different densities. From the comparison with the
respective Fig.~1 from the experiment in Ref.~\onlinecite{kellogg}
it is seen that our results reproduce very well all the
experimental findings for the densities $n=5.2,~4.7$ and $3.8\cdot
10^{10}$ cm$^{-2}$. For lower densities $n=1.7,~2.3$ and $3.1\cdot
10^{10}$ our calculations start to deviate from the experimental
results. This is clearly seen also in Fig.~\ref{lglg} where we
show a log-log plot of the drag resistivity versus density $n$ for
three different temperatures: $T=1,~2$, and $4$ K. The symbols are
our calculations while straight solid line represents $n^{-4}$
dependence. This figure corresponds to Fig.~2 from
Ref.~\onlinecite{kellogg} where it has been found that the
experimental results can be very well approximated by the quartic
density dependence. As seen in Fig.~\ref{lglg}, except very low
densities, our calculations of the drag resistivity, which include
the xc corrections, are also approximated by the quartic density
dependence quite well. This contrasts the RPA-based results where
the drag resistivity shows $n^{-3}$ dependence.

\section{Summary}

In this paper we have investigated the role of the many-body xc
effect on the dispersion of collective excitations and on the
frictional Coulomb drag in low density bi-layer 2DES. We have
calculated the spectrum of the bi-layer plasmons and the
temperature dependence of the drag resistivity in a broad range of
temperatures. Our calculations include both intra- and inter-layer
xc effects. These effects are critical in low density bi-layers
since the inter-layer separation in these structures is comparable
with the inter-particle spacing in an individual layer. We have
proposed a new approach and employed the full dynamic xc kernels
of the bi-layer 2DES in calculations of the acoustical and optical
plasmon dispersions as well as in the drag, mediated by exchange
of bi-layer plasmons. In studying the many-body effects on the
particle-hole contribution to drag we have still used the static
exchange-controlled LFF. The combined effect of the dynamic and
static many-body xc on drag has been investigated in comparison
with the RPA-based calculations of the transresistivity vs
temperature for different values of the electron density. At low
temperatures we have compared our results with the experimental
findings from Ref.~\onlinecite{kellogg}.

We have observed that after the inclusion of the full dynamic xc
kernels, a decrease of the electron density induces shifts of the
plasmon branches in opposite directions. And this is in stark
contrast to the tendency obtained within the RPA that both optical
and acoustical plasmons move away from the boundary of the PHC
with a decrease of the electron density. In the presence of the xc
effects, the optical mode still moves away from the PHC while the
acoustical branch moves closer to the PHC. The effect is so strong
that at the total $r_{s}=3$ the acoustical branch enters the PHC
and becomes completely destroyed at finite temperatures. These
changes of the plasmon spectrum, induced by the intra- and
inter-layer dynamic xc, are reflected in the temperature
dependence of the plasmon-mediated drag.

After inclusion of the xc effects the upturn temperature, which is
determined by the thermal excitations of the acoustical modes,
exhibits a "redshift" while the plasmon enhancement peak, which is
mainly due to the thermal excitation of high energy optical modes,
undergoes to a "blueshift" with decreasing carrier concentration.
Already for the total $r_{s}=2$ the upturn temperature decreases
so strongly that the bilayer plasmons heavily contributes to the
drag starting at temperatures less than $0.1T_{\text{F}}$, which
is approximately a factor of $2$ small than the upturn temperature
calculated within the RPA. The effect becomes even stronger for
the total $r_{s}=3$ so that the upturn temperature disappears
practically. Our results show that at low temperatures the dynamic
xc effects enhance the plasmon-mediated drag while at high
temperatures they have the opposite effect. We ascribe this to the
enhancement of the damping of high energy plasmons, caused by
dynamic many-body xc effects.

The combined effect of the dynamic and the static many-body xc in
low density bi-layers results in both quantitative and qualitative
changes in the behavior of the drag resistivity. At temperatures
near $0.5T_{F}$ the transresistivity with the xc effects included
is larger approximately by a factor of $3$ than that obtained
within the RPA for the in-layer $r^{in-layer}_{s}=2\sqrt{2}$. The
enhancement increases strongly at low temperatures where the drag
is dominated by large-angle scattering processes with $q\simeq
2k_{F}$. We have found that the plasmon enhancement peak, which is
clearly seen in the RPA based calculations, disappears at large
$r_s$ after the full xc corrections have been introduced. The
disappearance of the large high-temperature plasmon peak is the
result of the strong enhancement of the particle-hole contribution
to drag at low temperatures and of the moderate suppression of the
plasmon-mediated drag at high temperatures when, respectively, the
full static and dynamic xc corrections are included. We have
ascribed this behavior to the change in the specific contributions
to drag made by small- and large-angle scattering events. Our
calculations demonstrate clearly that the xc effects increase the
large-angle scattering component of the drag making it the
dominant contribution. This component strongly enhances the drag
for two reasons. First, at low temperatures the e-e scattering
phase-space diverges near $x=q/2k_{F}\simeq 1$. On the other hand,
when $x\simeq 1$, the static intra-layer LFF $G_{11}(x)$ becomes
close to unity, leading to a reduction of the effective
intra-layer interaction $V_{\text{eff},11}(x)$. This by itself
weakens the dynamic intra-layer screening and enhances the drag.
Thus, against the background of the large transresistivity at
small $T$, the plasmon-mediated contribution to drag does not
result in a peaked behavior of the scaled transresistivity and the
total drag rate decreases monotonically in $T$. In the RPA, the
drag still contains a strong small-angle scattering component
which suppresses partially the enhancement at small $T$ and the
plasmon peak remains visible.

Currently, the available experimental data for the low density
electron samples are restricted to low temperatures where the
plasmon contribution to drag is small. An experimental
verification of the plasmon peak disappearance requires new drag
measurements in low density electron bi-layers in a broad range of
temperatures close to $T_{F}$.

At low temperatures $T\ll T_{F}$ our numerical results are in good
agreement with the experimental findings by Kellogg \textit{et
al.}~\cite{kellogg}. At the upper edge of temperature interval
where the measurements are available our calculations of the drag
resistivity give about $90\%$ of the measured drag. This agreement
is especially important against the usual discrepancy between
theory and experiment in determination of the drag absolute value.
Our calculations beyond the RPA show that the drag resistivity as
a function of density exhibit $n^{-4}$ dependence for not very low
densities. This is in contrast to the RPA based calculations of
the drag resistivity and in agreement with the experimental
findings.

\section*{Acknowledgments}

We thank Zhixin Qian for useful discussions and for kindly making
the results of his calculations accessible prior to publication.
This work was supported by KOSEF Grant No.~R05-2003-000-11432-0
and NSF Grant No. DMR-0313681.

\end{document}